# Data-driven discovery of high performance layered van der Waals piezoelectric $NbOI_2$

Yaze Wu[1,2,9], Ibrahim Abdelwahab[2,5,9], Ki Chang Kwon[5], Ivan Verzhbitskiy[1,2], Lin Wang[5], Weng Heng Liew[6], Kui Yao[6], Goki Eda[1,2], Kian Ping Loh[2,5,7*], Lei Shen[3,4*], Su Ying Quek[1,2,7,8*]

[1]Department of Physics, National University of Singapore, Singapore, Singapore.

[2]Centre for Advanced 2D Materials and Graphene Research Centre, Singapore, Singapore.

[3]Department of Mechanical Engineering, National University of Singapore, Singapore, Singapore.

[4]Engineering Science Programme, National University of Singapore, Singapore, Singapore.

[5]Department of Chemistry, National University of Singapore, Singapore, Singapore.

[6]Institute of Materials Research and Engineering, A*STAR (Agency for Science, Technology and Research), Singapore.

[7]NUS Graduate School, Integrative Sciences and Engineering Programme, National University of Singapore, Singapore.

[8]Department of Materials Science and Engineering, National University of Singapore, Singapore.

[9]These authors contributed equally: Yaze Wu, Ibrahim Abdelwahab.

*e-mail: chmlohkp@nus.edu.sg; shenlei@nus.edu.sg; phyqsy@nus.edu.sg

**Abstract**

Using high-throughput first-principles calculations to search for layered van der Waals materials with the largest piezoelectric stress coefficients, we discover $NbOI_2$ to be the one among 2940 monolayers screened. The piezoelectric performance of $NbOI_2$ is independent of thickness, and its electromechanical coupling factor of near unity is a hallmark of optimal interconversion between electrical and mechanical energy. Laser scanning vibrometer studies on bulk and few-layer $NbOI_2$ crystals verify their huge piezoelectric responses, which exceed internal references such as $In_2Se_3$ and $CuInP_2S_6$. Furthermore, we provide insights into the atomic origins of anti-correlated piezoelectric and ferroelectric responses in $NbOX_2$ (X = Cl, Br, I), based on bond covalency and structural distortions in these materials. Our discovery that $NbOI_2$ has the largest piezoelectric stress coefficients among 2D materials calls for the development of $NbOI_2$-based flexible nanoscale piezoelectric devices.

Piezoelectric materials enable the interconversion between mechanical and electrical energy. This is made possible by the change in polarization of the material when it is stretched or compressed. As such, piezoelectric materials are integral components of intelligent, multi-functional devices and drive a multi-billion dollar industry[1] through their applications as sensors, actuators, energy harvesters, *etc.*[2-7]. The recent thrust toward flexible nanoscale devices creates a need for two-dimensional (2D) piezoelectric materials. Piezoelectric materials comprised of one or few layers of layered van der Waals (vdW) systems are particularly useful for increasingly important niche applications such as actuators with extreme atomic-scale precision[8] as well as wearable sensors and smart material applications that require a large voltage signal in response to a small amount of physical deformation. 2D piezoelectric materials provide a practical alternative to micro-scale battery packs, functioning as nano-generators to power nanoscale devices[9].

Thus far, the discovery of 2D piezoelectric materials has mostly been *ad hoc*, for example, by performing calculations on specific 2D materials that are known to be ferroelectric. However, with an *ad hoc* approach, it is difficult to ascertain if the 2D material indeed has optimal piezoelectric coefficients. Experimentally, it is also challenging to quantitatively compare the piezoelectric coefficients of 2D materials[10]. The objective of this work is to perform a systematic high throughput search through a 2D material database, in order to rank the 2D materials according to the size of their intrinsic piezoelectric coefficients. While 2D materials down to nanometres in thickness are sufficient for flexible nanoscale devices, symmetry-breaking in the monolayer can lead to the emergence of piezoelectricity in the monolayer even when the parent bulk materials are not piezoelectric[11]. Thus, we focus our search on monolayers. Out of 109 piezoelectric monolayers that we identify, the family of niobium oxydihalides $NbOX_2$ (X = Cl, Br, I) is predicted to have among the largest in-plane piezoelectric stress coefficients, an order of magnitude larger than those of most reported 2D materials. We note that $NbOX_2$ has recently been independently identified to be a robust room temperature ferroelectric in another high-throughput study searching for 2D ferroelectric materials[12]. While all ferroelectric materials are piezoelectric, there is no direct correlation between the magnitude of spontaneous polarization $|\vec{P}|$ and the magnitude of piezoelectric coefficients (see **Supplementary Fig. S1**). Within the $NbOX_2$ family, our calculations in fact show that the piezoelectric and ferroelectric effects have opposing trends down the halogen group. We further show that the large piezoelectric effect is independent of crystal thickness, in contrast to $MoS_2$ and similar 2D in-plane piezoelectrics, where the piezoelectricity vanishes for an even number of layers[13-15]. This thickness-independent piezoelectric effect is a practical advantage in isolating 2D nanoscale piezoelectrics. Experimental validations of the piezoelectric effect were carried out on few-layer $NbOI_2$ and $NbOCl_2$ crystals, where significantly larger piezoelectric coefficients were obtained compared to internal references such as $In_2Se_3$ and $CuInP_2S_6$ (known 2D piezoelectrics)[16-20]. Our findings pave the way for the development of $NbOI_2$-based flexible nanoscale piezoelectric devices, such as high precision actuators and wearable electronics or energy-harvesters.

## Results and discussion

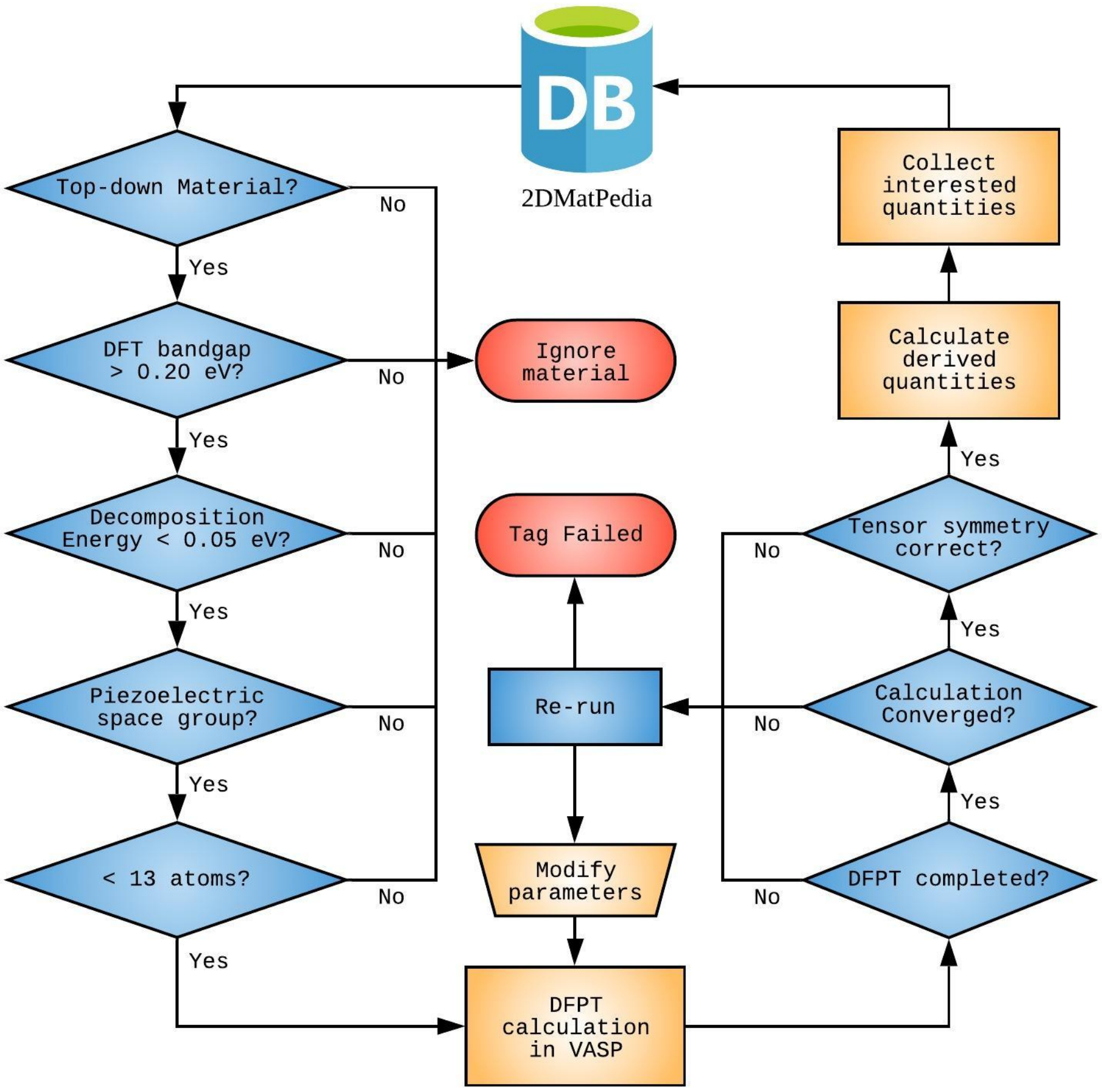

**Fig. 1 | Workflow of the high-throughput calculation to screen for piezoelectric 2D materials.** We have chosen a criteria of at least 0.20 eV for the DFT band gap (large enough for operation at finite temperatures) and a decomposition energy[21] of < 0.05 eV to ensure thermodynamic stability.

The workflow of our high-throughput calculations is shown in **Fig. 1**. Our results are publicly available in 2DMatpedia[21], an open database of 2D materials that shares the same infrastructure and basic workflow as the Materials Project database[22]. We focus only on the 2940 "top-down" materials within the database, which are obtained by exfoliation of known bulk layered materials, and are more likely to be dynamically stable and experimentally available. Next, we perform a rapid screening process based on the band gap and decomposition energy (both documented in 2DMatpedia[21]) as well as the space group (piezoelectric space groups lack inversion symmetry). A total of 225 materials pass this screening process. We then limit our high-throughput density functional perturbation theory (DFPT) calculations to materials with less than 13 atoms per unit cell (160 of the 225 materials). Following conventions for 2D materials, we compute the sheet piezoelectric stress tensor elements, $\boldsymbol{e}_{ij}$, defined as $\frac{\partial \boldsymbol{P}_i}{\partial \boldsymbol{\eta}_j} L$, the rate of change in polarization $\boldsymbol{P}_i$ with homogeneous strain $\boldsymbol{\eta}_j$ multiplied by the cell height $L$[23]. The index $i$ runs from 1 to 3 ($x$, $y$, $z$) and $j$ ranges from 1 to 6 ($xx$, $yy$, $zz$, $yz$, $xz$, $xy$) where the Voigt notation is used. A series of automated checks and analyses is carried out and the relevant data is saved into the database. 51 materials did not pass the automated checks. This is similar in proportion to those in other high throughput studies[1, 24, 25]; these materials were not studied in detail. The dynamical stability of individual materials is checked manually as needed outside this workflow.

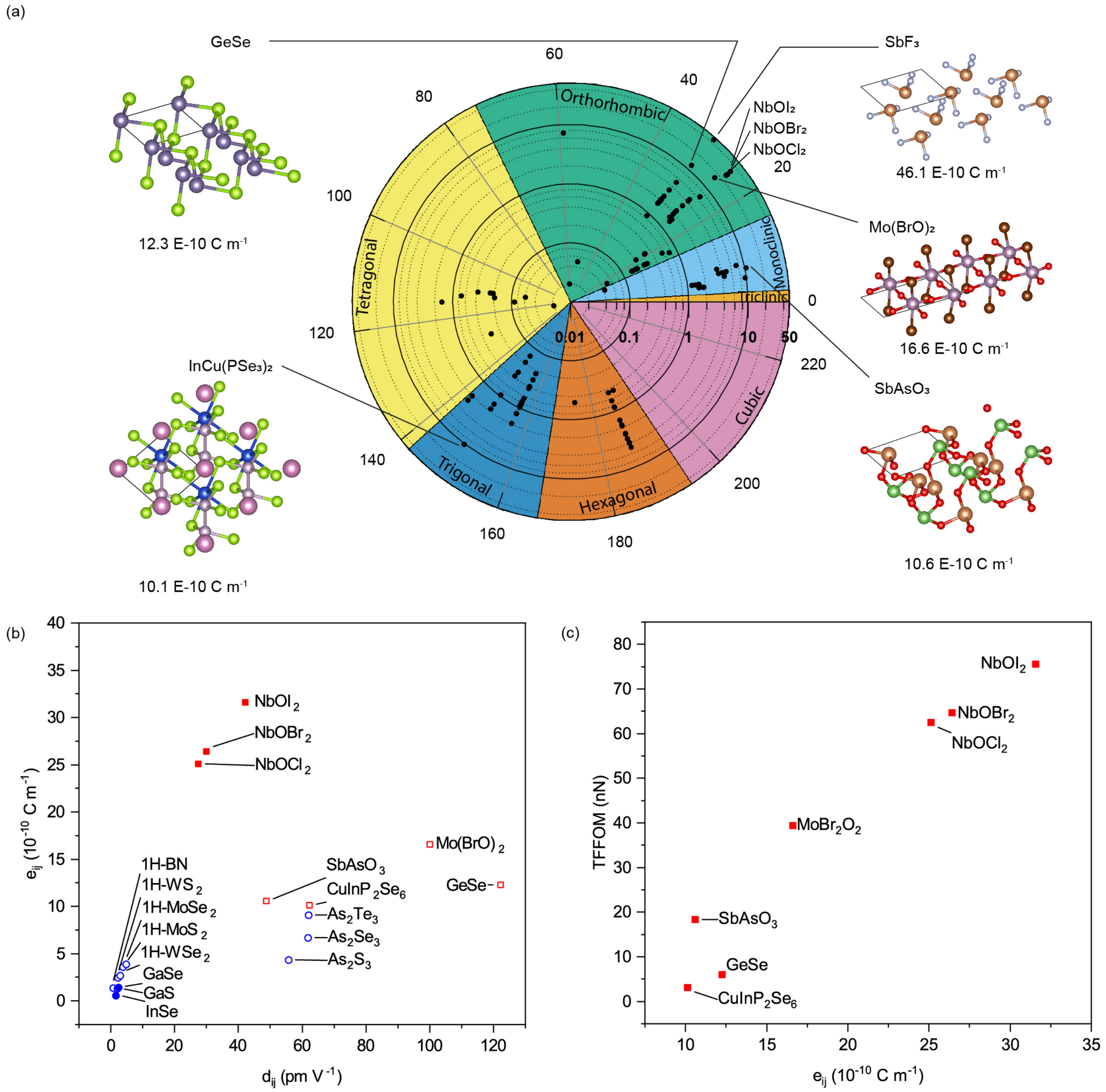

**Fig. 2 | Maximum sheet piezoelectric tensor elements, thin film figure of merit (TFFOM) and spontaneous polarizations. (a)** High-throughput calculation results for maximum sheet piezoelectric stress tensor elements ($\boldsymbol{e}_{ij}$). The radial axis represents the magnitude of $\boldsymbol{e}_{ij}$ in units of $10^{-10}\ C\ m^{-1}$ on a log scale and the angular axis represents the 230 space groups. Materials with $\boldsymbol{e}_{ij}$ larger than $10 \times 10^{-10}\ C\ m^{-1}$ are labeled and their atomic structures are presented. The maximum $e_{ij}$ corresponds to $\boldsymbol{e}_{26}$ for $SbF_3$ and to $\boldsymbol{e}_{11}$ for $NbOX_2$. **(b)** Plot of maximum sheet piezoelectric stress tensor elements ($\boldsymbol{e}_{ij}$) and

corresponding piezoelectric strain tensor elements ($\boldsymbol{d}_{ij}$) for various 2D materials. For additional notes on GeSe, please see **Supplementary Table S1**. **(c)** Plot of TFFOM and maximum $\boldsymbol{e}_{ij}$ of materials highlighted in (a). In (b), solid symbols denote materials that are piezoelectric in the thermodynamically most stable bulk form, as documented in the Materials Project database[1]; hollow symbols denote those that are not. Blue dots denote data points obtained from other studies[17, 23, 26-28] listed in **Supplementary Table S2**. Our computed $\boldsymbol{e}_{ij}$ values for these materials are also provided in **Supplementary Table S2** for comparison.

The final results are summarized in **Fig. 2a**. All the 109 materials (see **Supplementary Table S15** for the full list) have an $\boldsymbol{e}_{ij}$ value greater than $0.05 \times 10^{-10}$ C m$^{-1}$. These 109 2D piezoelectric materials are observed to belong to only a few space groups (**Fig. 2a**). Space groups 17, 26, 31, 149, 156 and 187 each have more than nine 2D piezoelectric materials. Our high throughput calculations also found that 48 of the 109 2D piezoelectric materials are not ferroelectric (see Supplementary Table S15). 2D materials previously identified to be piezoelectric[23, 26] are also found to be piezoelectric in our calculations, with values of $\boldsymbol{e}_{ij}$ very close to their reported values (**Supplementary Table S2**). Most materials have maximum $\boldsymbol{e}_{ij}$ values below $5 \times 10^{-10}\,\mathrm{C\,m^{-1}}$, while a few have significantly larger $\boldsymbol{e}_{ij}$. These larger $\boldsymbol{e}_{ij}$ values correspond to in-plane piezoelectricity.

We identify eight materials with maximum sheet $\boldsymbol{e}_{ij}$ values larger than $10 \times 10^{-10}\ C\ m^{-1}$, namely, $SbF_3$, $NbOI_2$, $NbOBr_2$, $NbOCl_2$, $MoBr_2O_2$, GeSe, $SbAsO_3$, $CuInP_2Se_6$. The structures of these eight materials (referred to as the 'top 8' materials) are presented in **Fig. 2a** and **Fig. 3**. $SbF_3$, which has the largest sheet $\boldsymbol{e}_{ij}$, is found to be dynamically unstable in the monolayer form. Niobium oxydihalides, $NbOX_2$, with X = Cl, Br, or I, have the next highest sheet $\boldsymbol{e}_{ij}$, with $NbOI_2$ having the largest $\boldsymbol{e}_{11}$ value of $\sim 32 \times 10^{-10}$ C m$^{-1}$. We have verified explicitly that monolayer $NbOI_2$ is dynamically stable (**Supplementary Fig. S2 and Fig. S3**) and we expect the same to be true for the other members of the family. The exfoliation energy for $NbOI_2$ monolayers is 18.2 meV Å$^{-2}$, lower than that of graphene (25.5 meV Å$^{-2}$)[21].

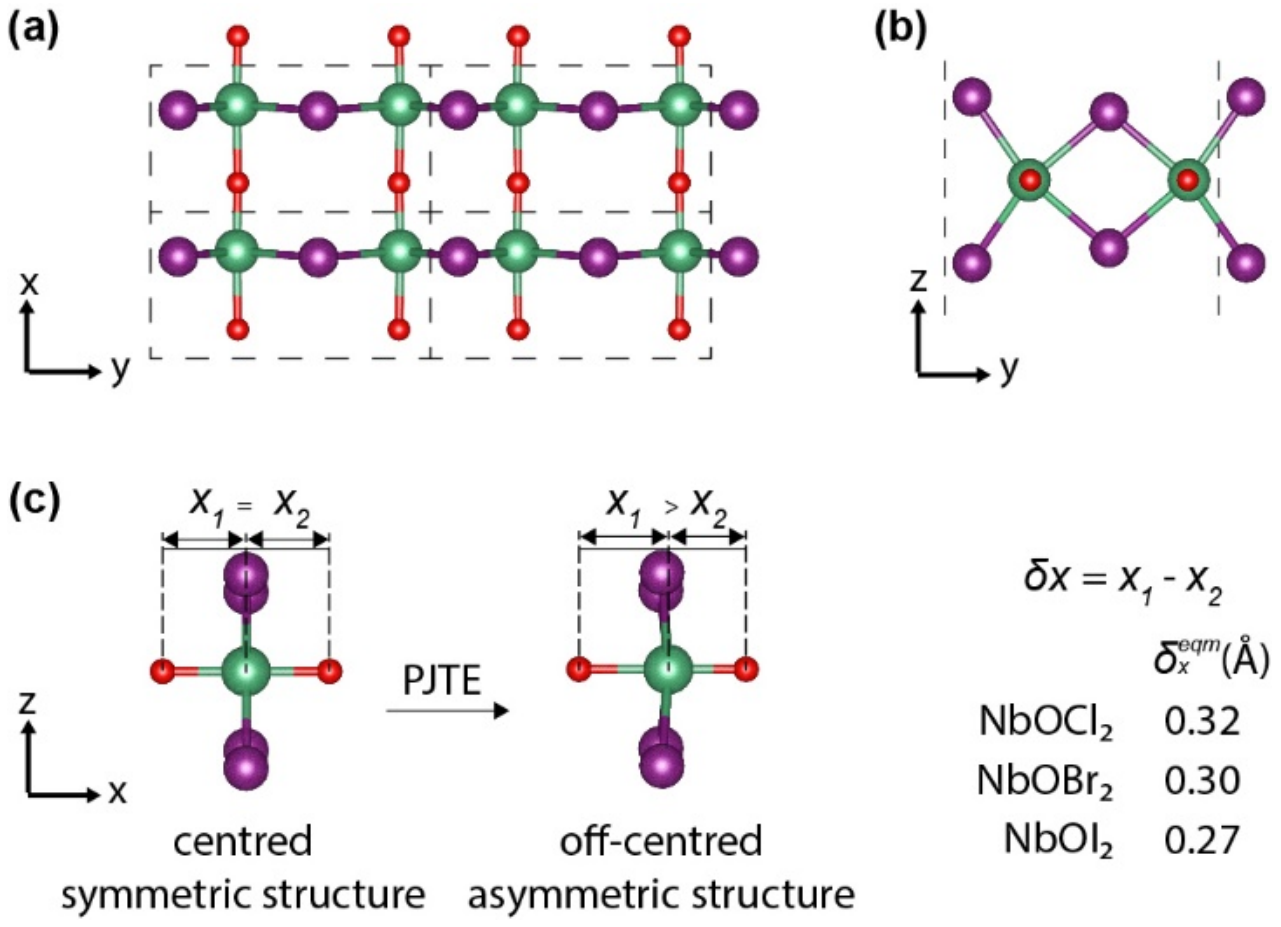

**Fig. 3 |** Atomic structure of monolayer $NbOX_2$ (X = I, Br, Cl). Dark green balls denote Nb atoms; purple balls denote halogens; red balls denote O atoms. Dashed lines mark the unit cell boundaries. (a) Top view. (b) Side view down the $x$-axis. (c) Schematic, viewed along $y$-axis, showing relative displacement of Nb atoms along the $x$-axis away from the high symmetry position where $\delta x = 0$ Å, to the equilibrium position where $\delta x = \delta_x^{eqm}$. Additional structural parameters are provided in **Supplementary Table S3**. PJTE refers to the pseudo-Jahn-Teller effect.

Besides the piezoelectric stress tensor elements $\boldsymbol{e}_{ij}$, the piezoelectric strain tensor elements $\boldsymbol{d}_{ij}$ are also widely discussed in the literature. The $\boldsymbol{d}_{ij}$ values are defined as $\frac{\partial \boldsymbol{P}_i}{\partial \boldsymbol{\sigma}_j}$, the rate of change in polarization $\boldsymbol{P}_i$ with homogeneous stress $\boldsymbol{\sigma}_j$ (see SI), and are related to $\boldsymbol{e}_{ij}$ according to $\boldsymbol{d}_{ij} = \boldsymbol{e}_{ik}\boldsymbol{S}_{kj}$, where $\boldsymbol{S}_{kj}$ is the compliance tensor (the inverse of the elastic tensor). Unlike $\boldsymbol{e}_{ij}$ and $\boldsymbol{S}_{kj}$ which have a different definition for the 2D case, the $\boldsymbol{d}_{ij}$ values are defined in the same manner for both bulk and 2D (see Formalisms in SI). The $\boldsymbol{d}_{11}$ values for monolayer $NbOX_2$ are ~42, 30 and 27 $pm\ V^{-1}$ for $NbOI_2$, $NbOBr_2$ and $NbOCl_2$, respectively (see **Supplementary Table S4**). We note that all the $\boldsymbol{d}_{ij}$ values for bulk $NbOX_2$ are almost the same as for the monolayer (**Table 2** and **Supplementary Table S5**), indicating that the piezoelectric properties of $NbOX_2$ are very similar from monolayer to bulk form. In the bulk material, the ratio of mechanical stress energy density to the electrical energy density is given by a dimensionless number $k^2 = \frac{\boldsymbol{e}_{ij}\boldsymbol{d}_{ij}}{\boldsymbol{\varepsilon}_{ij}\varepsilon_0}$ [29, 30], where $k$ is known as the electromechanical coupling factor. We obtain an electromechanical coupling factor of ~1.0, 0.9 and 0.9 for bulk $NbOI_2$, $NbOBr_2$ and $NbOCl_2$, respectively (see **Supplementary Table S5**). Since the maximum value of $k$ is unity, we see that the intrinsic piezoelectricity in $NbOX_2$ provides for highly efficient interconversion between electrical and mechanical energy. Our many-body perturbation theory calculations[31] predict a fundamental band gap of ~2.2 eV for monolayer $NbOI_2$, which is large enough for room temperature applications.

**Table 1 | Materials with $\boldsymbol{e}_{ij} > 10 \times 10^{-10}\ \boldsymbol{C\ m^{-1}}$.** $\boldsymbol{e}_{ij}$ is the piezoelectric stress tensor element, $\boldsymbol{d}_{ij}$ is the piezoelectric strain tensor element, $\boldsymbol{C}_{ij}$ is the elastic tensor element, and TFFOM is the figure of merit for thin film piezoelectric devices.[29] Our computed in-plane dielectric constant of $NbOX_2$ is 12 – 15 (see **Supplementary Table S6**).

| Material | ij | $\boldsymbol{e}_{ij}$ ($10^{-10}$ C m$^{-1}$) | $\boldsymbol{d}_{ij}$ ($pm\ V^{-1}$) | $\boldsymbol{C}_{ij}$ ($N\ m^{-1}$) | TFFOM ($nN$) |
|---|---|---|---|---|---|
| $NbOI_2$ | 11 | 31.6 | 42.2 | 75.6 | 71.7 |
| $NbOBr_2$ | 11 | 26.4 | 30.0 | 89.0 | 63.2 |
| $NbOCl_2$ | 11 | 25.1 | 27.4 | 92.9 | 59.6 |
| $MoBr_2O_2$ | 22 | 16.6 | 100.1 | 33.8 | 39.3 |
| GeSe | 22 | 12.3 | 122.2 | 18.2 | 6.0 |
| $SbAsO_3$ | 11 | 10.6 | 48.7 | 22.3 | 18.4 |
| $CuInP_2Se_6$ | 22 | 10.1 | 62.3 | 44.5 | 3.1 |

**Table 2 | Piezoelectric tensor elements and spontaneous polarizations of $NbOX_2$.** Sheet piezoelectric stress tensor elements $\boldsymbol{e}_{ij}$ are in units of $10^{-10}$ C m$^{-1}$ and piezoelectric strain tensor elements $\boldsymbol{d}_{ij}$ are in pm V$^{-1}$. Piezoelectric tensor elements that are zero due to symmetry of the space group are omitted here. Spontaneous polarization along the $x$-direction ($\boldsymbol{P}_x$) in pC m$^{-1}$ is calculated as the difference between the polarization of the equilibrium and symmetric structures.

| Formula | $\boldsymbol{e}_{11}$ | $\boldsymbol{e}_{12}$ | $\boldsymbol{e}_{13}$ | $\boldsymbol{e}_{26}$ | $\boldsymbol{e}_{35}$ | $\boldsymbol{d}_{11}$ | $\boldsymbol{d}_{12}$ | $\boldsymbol{d}_{26}$ | $\boldsymbol{P}_x$ |
|---|---|---|---|---|---|---|---|---|---|
| $NbOCl_2$ | 25.1 | -1.1 | -0.4 | 0.8 | 0.0 | 27.4 | -4.1 | 5.4 | 185 |
| $NbOBr_2$ | 26.4 | -1.0 | -0.4 | 0.8 | 0.0 | 30.0 | -4.1 | 5.8 | 170 |
| $NbOI_2$ | 31.6 | -1.0 | -0.3 | 0.7 | 0.0 | 42.2 | -5.1 | 5.2 | 143 |

We compute the $\boldsymbol{d}_{ij}$ values for the top 8 materials (excluding $SbF_3$) and the largest components of $\boldsymbol{e}_{ij}$ and $\boldsymbol{d}_{ij}$ are shown in **Fig. 2b** and **Table 1**. The maximum $\boldsymbol{e}_{ij}$ and $\boldsymbol{d}_{ij}$ values for other 2D piezoelectrics discussed in the literature are also plotted in **Fig. 2b** for comparison (see also **Supplementary Table S2**). Some materials such as monolayer GeSe[27], $As_2S_3$[17], $As_2Se_3$[17] have large $\boldsymbol{d}_{ij}$ values but small $\boldsymbol{e}_{ij}$ values, corresponding to small values of their Young's moduli. The small Young's moduli limits the amount of force exerted in electric field-induced deformations. (The 2D Young's moduli $C_{11}$ for monolayer $NbOI_2$ is ~76 N/m (**Supplementary Table S7**) while the bulk value for $C_{11}$ is ~125 GPa.) A figure of merit adopted for thin-film piezoelectrics (TFFOM), when the passive elastic layer is much thicker than the piezoelectric material, is $\frac{\boldsymbol{e}_{ij}^2}{\boldsymbol{\varepsilon}_{ij}\boldsymbol{\varepsilon}_0}$, where $\boldsymbol{\varepsilon}_{ij}$ is the dielectric constant and $\boldsymbol{\varepsilon}_0$ is the vacuum permittivity[29, 32, 33]. Thus, the piezoelectric stress coefficients rather than the piezoelectric strain coefficients are particularly important for 2D flexible piezoelectric applications.

It is clear that $NbOX_2$ has the largest $\boldsymbol{e}_{ij}$ values in **Fig. 2b**. To our knowledge, there is only one 2D material, SnSe, with a predicted $\boldsymbol{e}_{11}$ value[27] that is larger than $NbOI_2$. Recent experiments on SnSe have, however, reported a much weaker piezoelectric performance[9] (see **Supplementary Table S8**), which we attribute to the Poisson effect, which reduces the effective $\boldsymbol{e}_{11}$ value of SnSe by ~51% (see **Supplementary Table S1**). We note that our high throughput calculations do not account for the Poisson effect. However, the Poisson effect does not change the $\boldsymbol{e}_{11}$ values for $NbOX_2$ (see **Supplementary Table S1**). The figure of merit (TFFOM) for our top 8 candidates from the high throughput calculations (except $SbF_3$) are presented in **Fig. 2c** and **Table 1**, where it is clear that $NbOI_2$ has the largest TFFOM.

In **Fig. 2b**, we also indicate using solid symbols the materials that are piezoelectric in the thermodynamically most stable bulk form, and hollow symbols for those that are not. $NbOX_2$ are among the minority of 2D materials that are piezoelectric both in the monolayer and in the bulk. The $\boldsymbol{d}_{ij}$ values are essentially the same for both monolayer and bulk $NbOX_2$ (see **Table S9**). Of the 109 piezoelectric materials that we discovered, only 30 are also piezoelectric in the bulk, according to a similar high throughput study[1] on the Materials Project database for bulk materials, from which the candidate monolayers were derived. A comparison of the largest $\boldsymbol{e}_{ij}$ values from the two independent studies indicates that the piezoelectric coefficients in the monolayer and the bulk are strongly correlated (see Supplementary **Table S10** and Supplementary **Fig. S4**). The thickness-independent piezoelectric effect in $NbOX_2$ implies that few-layer $NbOX_2$ can be prepared for nanoscale piezoelectric applications without the need for a pre-selection process based on the number of layers.

The atomic structure of monolayer $NbOX_2$ is shown in **Figs. 3a** and **3b**. Along the $y$-direction, Peierls distortion results in alternating Nb-Nb distances, and along the $x$-direction, the Nb atom is displaced away from the high symmetry position where $\delta x = x_1 - x_2$ (**Fig. 3c**), giving rise to a spontaneous polarization (**Table 2**). The degree of structural asymmetry along the $x$-direction is largest for $NbOCl_2$ and smallest for $NbOI_2$ (**Fig. 3c**). The asymmetry along the $x$-direction can be explained by the pseudo-Jahn-Teller effect (PJTE)[34], where mixing between the valence O $p$ orbitals and conduction Nb $d$ orbitals results in a more energetically favourable configuration accompanied by structural distortion as well as increased Nb-O bond strengths and covalency (see SI for details). The piezoelectric tensor elements in **Table 2** reflect the strong in-plane anisotropy of the system. In bulk $NbOX_2$, the directions for the inversion-symmetry-breaking distortions are the same in all layers, and the difference in $\delta_x^{eqm}$ is within 0.002 Å in the bulk and monolayer systems (**Supplementary Table S3** and **Table S11**). Thus, the piezoelectric properties are similar in bulk, monolayer and thin film form.

Quantitative measurements of piezoelectric coefficients in 2D materials are challenging, since any small parasitic vibration, boundary effect, or electrostatic force during the piezoelectric measurement (especially the single-point measurement) significantly affect the accuracy of the measured values[10]. To provide quantitative information about the piezoelectric coefficients within the family of $NbOX_2$, and to compare the piezoelectric coefficients with those of other known 2D piezoelectrics, $\alpha$-$In_2Se_3$ and $CuInP_2S_6$, we performed laser scanning vibrometer (LSV)[35] measurements on thick bulk-like samples. All these materials are also piezoelectric in the bulk. The existence of ferroelectricity and piezoelectricity in thin films of $NbOX_2$ was demonstrated using piezoresponse force microscopy (PFM) for thicknesses down to sub-10 nm.

We synthesize large-sized $NbOI_2$ and $NbOCl_2$ crystals grown by the chemical vapor transport method (refer to Methods for details). The crystal structures of the as-grown $NbOI_2$ and $NbOCl_2$ crystals are confirmed by single-crystal X-ray diffraction (SC-XRD) (**Supplementary CIFs** and **Supplementary Table S11**). The crystallographic directions of the $NbOX_2$ crystals are also identified from the SC-XRD analysis. These room temperature SC-XRD studies show that the crystal belongs to polar space group C2 (No. 5), hence providing further evidence that the ground state of bulk $NbOX_2$ is ferroelectric (refer to **Supplementary Note 1: Polarization switching in $NbOX_2$** and **Supplementary Note 2: Ferroelectric-paraelectric phase transition in $NbOI_2$** for ferroelectric data).

To investigate if ultrathin $NbOX_2$ is ferroelectric, nanosheets were exfoliated from bulk crystals via the Scotch tape method, and then transferred onto gold (Au) substrates (**Supplementary Fig. S5**) for PFM characterization. $NbOX_2$ is thermodynamically stable and all experimental measurements were performed under ambient conditions. In PFM, an AC voltage is applied between a conductive sharp tip and the bottom electrode of a piezoelectric sample to induce local mechanical deformations by means of the converse piezoelectric effect[36, 37]. Scanning tip-induced hysteretic switching events were recorded using spectroscopic PFM (**Supplementary Note 1: Polarization switching in $NbOX_2$,** and **Fig. S18**). The ferroelectric-paraelectric phase transition in $NbOI_2$ was further observed using temperature-dependent differential scanning calorimetry and second harmonic generation measurements (**Supplementary Note 2: Ferroelectric-paraelectric phase transition in $NbOI_2$,** and **Fig. S19**).

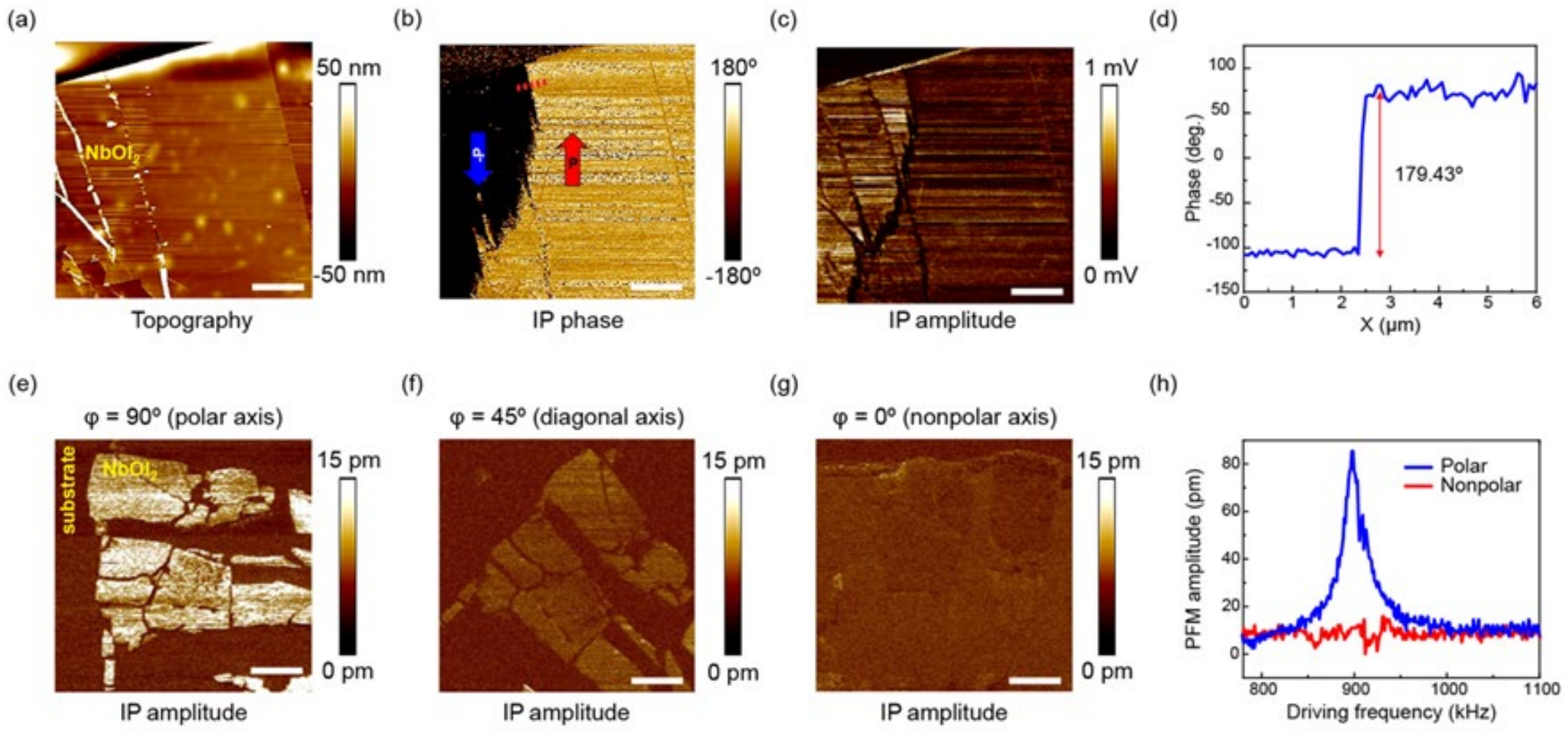

**Fig. 4 | Piezoelectric force microscopy (PFM) investigation of thin $NbOX_2$.** (**a**) Topography, (**b**) In-plane (IP) phase, (**c**) IP amplitude, and (**d**) phase profile across antiparallel polarization states of 82-nm-thick $NbOI_2$ flake. (**e-g**) Vector PFM IP amplitude images of 10-nm-thick $NbOI_2$ showing spontaneous polarization at 90° (e), 45° (f), and 0° (g) angles relative to the cantilever long axis. (**h**) PFM amplitude profiles along the polar and nonpolar axes of the 10-nm-thick $NbOI_2$ flake. Scale bars: 4 μm. Drive voltage: (b,c) 5 V, (e-h) 3 V. Drive frequency: (b,c) 65 kHz, (e) 827 kHz, (f) 974 kHz, (g) 890 kHz.

**Figs. 4b** and **4c** present phase and amplitude images constructed by in-plane (IP) PFM output channel from 82-nm-thick $NbOI_2$ flake (**Fig. 4a**). A clear contrast exists between different domains in the lateral phase and amplitude. The PFM phase indicates the direction of the ferroelectric polarization, whereas the PFM amplitude reflects the magnitude of the local piezoelectric response. The bright and dark contrasts in the PFM phase image indicate that there exist two oppositely polarized ferroelectric domains, characterized by a ~180° phase difference (**Fig. 4d**). **Supplementary Fig. S6-Fig. S10** display additional PFM data for $NbOI_2$ and $NbOCl_2$, demonstrating piezoelectric response down to 4.3 nm-thickness for $NbOI_2$.

To confirm the in-plane piezoelectric anisotropy, vector PFM[38, 39] is conducted on 10-nm-thick $NbOI_2$ (**Fig. 4e-h**) and 17-nm-thick $NbOCl_2$ (**Supplementary Fig. S11**) flakes with height profiles shown in **Supplementary Fig. S12**. As expected, the in-plane PFM response is strongest when the polar axis is orthogonal to the cantilever long axis, as shown in **Fig. 4e** (φ is the polarization angle relative to the cantilever long axis). When the sample is rotated by 90 degrees, no lateral PFM contrast is observed (**Fig. 4g, h**). In addition, the out-of-plane piezoresponses for the $NbOX_2$ nanoflakes are typically negligible, as displayed in **Supplementary Fig. S6-Fig. S10**. The vector PFM results as well as predominantly in-plane response confirms that the physical origin of the PFM signals is piezoelectricity rather than electrostatic tip-sample interactions[40, 41].

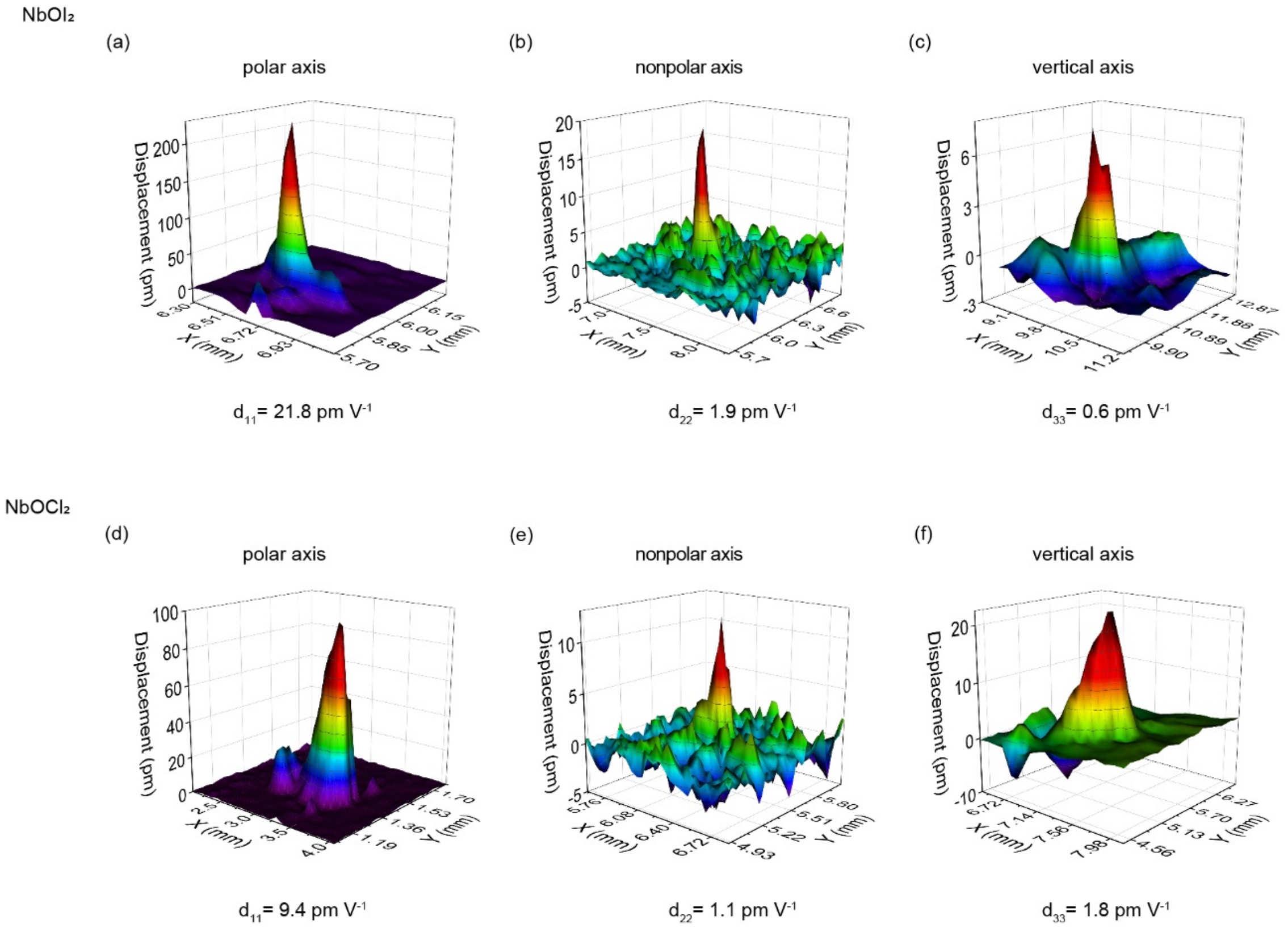

**Fig. 5 | Evaluation of the piezoelectric coefficients of $NbOX_2$ using a laser scanning vibrometer (LSV).** 3D graphs of the instantaneous vibration when the displacement magnitude reaches the maximum under the sine-wave driving electrical signal. **(a,d)** Measurement along the lateral polar direction ($\boldsymbol{d_{11}}$). **(b,e)** Measurements along the lateral nonpolar direction ($\boldsymbol{d_{22}}$). **(c,f)** Measurements along the vertical nonpolar direction ($\boldsymbol{d_{33}}$).

We quantified the lateral ($\boldsymbol{d_{11}}$, $\boldsymbol{d_{22}}$) and vertical ($\boldsymbol{d_{33}}$) piezoelectric coefficients of $NbOX_2$ through laser scanning vibrometer (LSV)[35] measurements. LSV is a non-contact optical technique that measures the vibration velocity of a moving piezoelectric surface by monitoring the interference pattern (Doppler frequency shift) between the scattered light and the incident light. By integrating the vibration velocity, the mechanical displacement in response to an applied electric field can be determined. Consistent with the vector PFM data, the LSV vibration modalities (Fig. 5) further confirm the anisotropy of the piezoelectric response in $NbOX_2$. Our measurements reveal that $NbOI_2$ exhibits stronger piezoelectric effects than $NbOCl_2$ (the $\boldsymbol{d}_{11(eff)}$ values are ~ 21.8 pm $V^{-1}$ for $NbOI_2$ and ~ 9.4 pm $V^{-1}$ for $NbOCl_2$), consistent with our theoretical predictions (**Table 2**). These values are likely to be underestimated due to the significant electric leakage of the $NbOX_2$ samples. Despite this, our measured $\boldsymbol{d}_{11(eff)}$ piezoelectric coefficients are larger than the largest $\boldsymbol{d}_{ij(eff)}$ values we measured for common 2D ferroelectrics, such as α-$In_2Se_3$ and $CuInP_2S_6$ (**Supplementary Fig. S13**). There is also a strong correlation between the measured $\boldsymbol{d}_{ij(eff)}$ values and the corresponding computed values (**Supplementary Fig. S14**). These LSV results show that $NbOX_2$ has superior piezoelectric performance compared to other 2D material piezoelectrics.

We analyse the origins of the large piezoelectric and ferroelectric effects in $NbOX_2$, and their trends down the halogen group. The large values of $\boldsymbol{e}_{11}$ arise predominantly from the lattice response to the applied strain (see **Supplementary Table S12**). This dominant ionic contribution can be written in an implied sum notation as

$$\boldsymbol{e}_{ij}^{ion} = \boldsymbol{Z}_{m,i}^{*} \frac{\partial \boldsymbol{u}_m}{\partial \boldsymbol{\eta}_j} \qquad 1$$

where $\boldsymbol{Z}_{m,i}^{*}$ is the dynamical charge ($m$ is a composite label for atom and displacement direction and $i$ is the direction of the polarization), $\boldsymbol{u}$ is the position vector of the atom and $\boldsymbol{\eta}_j$ denotes the applied strain. The dynamical charge is defined as the rate of change of polarization with atomic displacement, while $\frac{\partial \boldsymbol{u}_m}{\partial \boldsymbol{\eta}_j}$ quantifies the rate of change in atomic displacement with applied strain. The dynamical charges play an important role in both the piezoelectric and ferroelectric effects.

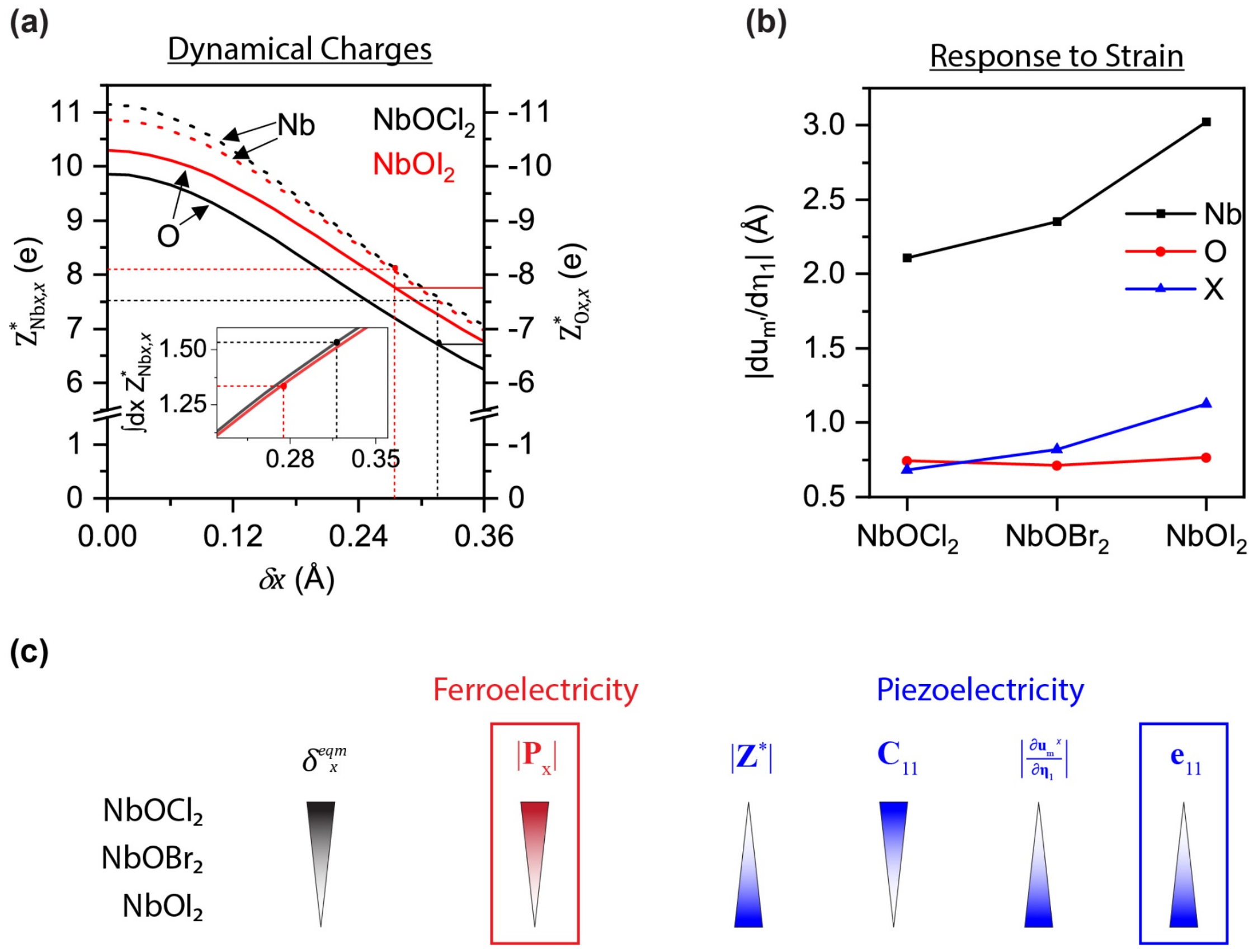

**Fig. 6 | Understanding the origin of piezoelectric and ferroelectric effects in the $NbOX_2$ family. (a)** Dynamical charges for Nb and O, $\boldsymbol{Z}^*_{Nbx,x}$ and $\boldsymbol{Z}^*_{Ox,x}$. as a function of $\delta x$, the difference between the longer and shorter Nb-O bonds. The inset shows the integral of $\boldsymbol{Z}^*_{Nbx,x}$ with respect to the $x$-displacement away from the high-symmetry site. In both the main figure and the inset, dashed lines indicate the values in the equilibrium structures. **(b)** $\left|\frac{\partial \boldsymbol{u}_m{}^x}{\partial \boldsymbol{\eta}_1}\right|$ in the equilibrium structures of $NbOX_2$, indicating the rate of change of $x$-displacement of each atom (Nb, O or X) with strain in the $x$-direction. $NbOI_2$ has the largest lattice response to strain, due to the smaller stiffness tensor elements and weaker Nb-O bonds. **(c)** Schematic illustrating the origins for the trends in ferroelectricity and piezoelectricity in $NbOX_2$. In (a) and (b), the respective values for $NbOBr_2$ are between those of $NbOCl_2$ and $NbOI_2$, and are not shown here.

Fig. 6a shows that the dynamical charges in the $x$-direction for Nb and O in $NbOCl_2$ and $NbOI_2$, decrease in magnitude as the off-centre displacement, $\delta x$, increases. These dynamical charges are significantly larger in magnitude than their expected formal oxidation states (+4 for Nb and -2 for O) as well as their estimated static charges (see Supplementary Fig. S15). The anomalous dynamical charges can be attributed to the partial covalency present in the Nb-O bonds, which is more significant in $NbOI_2$ due to the smaller electronegativity of I compared to Cl (see SI and Supplementary Fig. S16)[42]. The smaller off-centre distortion in $NbOI_2$ (smallest $\delta_x^{eqm}$) further increases the dynamical charges at equilibrium compared to those for $NbOCl_2$, contributing to the superior piezoelectric performance of $NbOI_2$. On the other hand, the integral of the dynamical charges with respect to atomic displacements, from the centred symmetric structure to the equilibrium off-centred structure, gives the magnitude of the spontaneous polarization at equilibrium, and the larger $\delta_x^{eqm}$ for $NbOCl_2$ results in a larger $|\vec{P}|$ compared to $NbOI_2$ (Fig. 6a inset).

The larger magnitudes of $\frac{\partial \boldsymbol{u}_m{}^x}{\partial \boldsymbol{\eta}_1}$ (the superscript for $m$ representing displacements in the $x$-direction) for Nb and X atoms in $NbOI_2$ (Fig. 6b) further contribute to the large $\boldsymbol{e}_{11}$ value for $NbOI_2$ (see also Supplementary Table S13). The observed trend in $\frac{\partial \boldsymbol{u}_m{}^x}{\partial \boldsymbol{\eta}_1}$ for Nb and X can be traced to the degree of bond covalency/ionicity in these systems, the more ionic $NbOCl_2$ having stiffer bonds and hence smaller magnitudes of $\frac{\partial \boldsymbol{u}_m}{\partial \boldsymbol{\eta}_j}$. This trend can also be observed in the decrease in Young's modulus $\boldsymbol{C}_{11}$ down the halogen group (see Supplementary Table S7). Fig. 6c summarizes the different factors that contribute to $\boldsymbol{e}_{11}$ being largest for $NbOI_2$, in contrast to $|\vec{P}|$ being largest for $NbOCl_2$.

While our top candidates for monolayer piezoelectrics all exhibit in-plane piezoelectricity, we comment that out-of-plane piezoelectricity (non-zero $\boldsymbol{e}_{3j}$) in the 2D monolayers is found in 46 of the materials in our database, as shown in **Supplementary Fig. S17**. Most of the non-zero $\boldsymbol{e}_{3j}$ values have magnitudes less than $1 \times 10^{-10}$ C m$^{-1}$, with a few exceptions having values up to ~$4 \times 10^{-10}$ C m$^{-1}$, as indicated in **Supplementary Fig. S17** and **Supplementary Table S14**. It has been reported that bulk layered $CuInP_2S_6$ has an electromechanical coupling factor of 0.7-0.9, corresponding to a large out-of-plane piezoelectric coefficient which originates from the deformation of the van der Waals gap[43]. Such a mechanism has not been considered in our study on monolayers. Our findings that monolayer and bulk $NbOI_2$ have exceptionally large in-plane piezoelectric effects are especially important for actuator applications that require pure in-plane movement[44].

In summary, we have identified, from among 2940 candidate monolayers, $NbOI_2$ as the material with the largest piezoelectric stress coefficient. Furthermore, $NbOI_2$ is one of the minority of monolayer piezoelectrics that is also piezoelectric in the bulk, with thickness-independent piezoelectric coefficients. Our experimentally measured values of piezoelectric strain coefficients are within a factor of two of the predicted value for $NbOI_2$, and are much larger than those measured concurrently for $\alpha$-$In_2Se_3$ and $CuInP_2S_6$. We have also verified in-plane ferroelectricity in $NbOI_2$. While $NbOI_2$ has the largest piezoelectric coefficients, $NbOCl_2$ has the largest spontaneous polarization. The excellent piezoelectric and ferroelectric effects in $NbOX_2$, as well as their trends down the halogen group, are rationalized on the basis of bond covalency and symmetry-breaking structural distortions in these materials. The structure-property correlations obtained here provide guidance for the design of functional piezoelectric and ferroelectric materials, while the discovery of 2D $NbOI_2$ as a high-performance piezoelectric paves the way for 2D piezoelectric devices.

## Methods

**Computational methods**. For the high-throughput DFPT calculations, only materials from space groups 1, 3-9, 16-46, 75-82, 89-122, 143-146, 149-161, 168-174, 177-190, 195-199, 207-220[1] are selected. These space groups lack inversion symmetry. DFPT calculations are performed with the plane-wave pseudopotential code VASP[45, 46], employing the generalized gradient approximation (GGA) for the exchange-correlation functional[47]. We use an energy cutoff of 520 eV, a Monkhorst-Pack k-point mesh with a density of 1500 per reciprocal atom (number of atoms per cell multiplied by the number of k-points)[1], a force convergence criterion of 0.005 eV/Å and a criterion of $10^{-10}$ eV for the convergence of the self-consistent cycle. The calculations are performed with a vacuum separation of about 20 Å between the 2D materials.

We modify the standard workflows[48, 49] developed by Materials Project[1, 22] to include additional post processing steps that convert the 3D piezoelectric tensor elements to 2D sheet piezoelectric tensor elements in the unit of C $m^{-1}$ by multiplying the former with the cell height. Other results, including the dynamical charge tensor (also known as the Born effective charge tensor), dielectric tensor (ionic and electronic contributions), Γ point phonon eigenvalue and eigenvectors, full piezoelectric tensor (in both C $m^{-2}$ and C $m^{-1}$), maximum (sheet) piezoelectric tensor elements in C $m^{-2}$ (C $m^{-1}$) and maximum (sheet) out-of-plane piezoelectric tensor elements in C $m^{-2}$ (C $m^{-1}$) are also captured. Our workflow also inherits the consistency checks and filters in the Materials Project Workflow[1] to detect errors arising from DFT calculation and convergence-related issues.

For the targeted study on $NbOX_2$, Monkhorst-Pack k point meshes of 12×6×1 including the Γ point are used. The atomic coordinates are fully relaxed using the conjugate gradient scheme until the maximum energy difference between iterations is less than $10^{-8}$ eV and the residual force is less than 0.001 eV/Å. Other parameters are inherited from the high-throughput calculations.

For calculations with fixed $\delta x$, the $x$-coordinates of the atomic positions are fixed while the $y$ and $z$ coordinates are allowed to relax. The lattice constants are kept fixed. $\frac{\partial \boldsymbol{u}_{\boldsymbol{m}}{}^{x}}{\partial \boldsymbol{\eta}_1}$ is obtained by applying strain in the $x$-direction and allowing the atoms to relax.

The elastic properties of $NbOX_2$ are obtained using a finite difference method as implemented in VASP. The elastic tensors are adapted for 2D materials according to the treatment by Choudhary et al.[50].

The piezoelectric strain tensor ($\boldsymbol{d}_{ij}$) is calculated from piezoelectric stress tensor ($\boldsymbol{e}_{ij}$) through **Supplementary Equation S7**. Elements in $\boldsymbol{S}_{ij}$ related to the $z$-direction are set to 0.

**Synthesis of single crystals of $NbOI_2$ and $NbOCl_2$.** Crystalline $NbOI_2$ and $NbOCl_2$ are grown by the chemical vapor transport method. High-purity Nb (film), iodine (crystals) and $Nb_2O_5$ (powder) with a stoichiometric ratio Nb:O:I = 1:1:2 are used as precursors for the growth of $NbOI_2$. The mixture of precursors is placed in a quartz ampule, and the ampule is sealed after being evacuated ($10^{-5}$ Torr). Similarly, Nb, $NbCl_5$ (powder) and $Nb_2O_5$ are used to grow $NbOCl_2$. The sealed quartz ampules are placed at the centre of a horizontal dual-zone furnace. Both heating zones were slowly heated to 600 °C and holding for 5 days. The ampules are then slowly cooled for 10 days with slightly different rates at the hot (1.2 °C/hour) and cold (1.5 °C/hour) zones. After the slow-cooling process, the furnace is turned off allowing the ampules to cool down naturally. Crystals are extracted from the opened ampules under inert conditions of an $N_2$-filled glove box and then stored for future use.

**X-ray diffraction (XRD) measurements**. Single crystal X-ray diffractions of bulk $NbOCl_2$ and $NbOI_2$ crystals are measured using a four circles goniometer Kappa geometry, Bruker AXS D8 Venture, equipped with a Photon 100 CMOS active pixel sensor detector. A molybdenum monochromatized (λ = 0.71073 Å) X-Ray radiation is used for the measurement. The frames are integrated with the Bruker SAINT software using a narrow-frame algorithm. Data is corrected for absorption effects using the Multi-Scan method (SADABS). The structures are solved in the monoclinic unit cell and refined using the SHELXT, VERSION 2014/5 Software. The final anisotropic refinement of the structures is performed by least squares procedures on weighted $F^2$ values using the SHELXL-2014/7 (Sheldrick, 2014) included in the APEX3 v2016, 9.0, AXS Bruker program.

**Sample preparation and piezoresponse force microscopy (PFM) characterization**. Mechanical exfoliation is performed by peeling off as-grown crystals using the Scotch tape method. Exfoliated crystals are directly transferred onto fresh gold (Au) substrates for PFM characterization. PFM images are obtained using a Bruker Dimension Icon AFM in contact mode. Pt/Ir-coated silicon tips with a radius of 20 nm and a force constant of 0.4 N/m are used for the PFM measurements. The drive frequency and drive amplitude for the PFM images are ~30-975 kHz and 2.5-10 V, respectively. The PFM amplitude is expressed in pm or mV units depending on the selected PFM output channel. Angular-resolved vector PFM was performed by rotating the $NbOX_2$ samples with respect to the cantilever axis.

**Laser scanning vibrometer (LSV) measurements**. The effective piezoelectric coefficients of $NbOX_2$, α-$In_2Se_3$, and $CuInP_2S_6$ (CIPS) are measured with a laser scanning vibrometer (OFV- 3001-SF6, PolyTech GmbH) after the crystals are DC poled under 150 V along the relevant crystallographic direction for 5 minutes. The LSV data are collected along the $x$ ($d_{11}$), $y$ ($d_{22}$) and $z$ ($d_{33}$) directions of $NbOX_2$ and along the $z$ ($d_{33}$) directions of

α-$In_2Se_3$ and CIPS under a unipolar AC signal of amplitude 10 V at 8 kHz through silver electrodes. The α-$In_2Se_3$ and $CuInP_2S_6$ crystals were purchased from HQGraphene. The effective piezoelectric coefficients are deduced from the profile analysis of the instantaneous displacement data to determine the strain generated under the sine-wave driving electrical signal.

**Differential scanning calorimetry (DSC)** analysis was performed under a nitrogen atmosphere with a heating rate of 10 °C/min using Mettler-Toledo DSC.

The polarization−electric field (P–E) curve was recorded using a ferroelectric tester (Precision Multiferroic II, Radiant Technologies).

Mechanical exfoliation is performed by peeling off as-grown crystals using the Scotch tape method. Exfoliated crystals are directly transferred onto fresh gold (Au) substrates for PFM characterization. PFM images are obtained using a Bruker Dimension Icon AFM in contact mode. Pt/Ir-coated silicon tips with a radius of 20 nm and a force constant of 0.4 N/m are used for the PFM measurements. The drive frequency and drive amplitude for the PFM images are ~30-975 kHz and 2.5-10 V, respectively. The PFM amplitude is expressed in pm or mV units depending on the selected PFM output channel. Angular-resolved vector PFM was performed by rotating the $NbOX_2$ samples with respect to the cantilever axis. The spectroscopic PFM hysteresis loops were acquired with ±10 V DC sweeps while applying an AC voltage of 5 V.

## Acknowledgements

Y.W. acknowledges support from the NUS Research Scholarship. K.Y and W.H.L. acknowledge partial support from the National Research Foundation, under the Competitive Research Programme of Singapore, NRF-CRP15-2015-04, and A*STAR, under RIE2020 AME Individual Research Grant (IRG) (Grant No.: A20E5c0086). G.E. acknowledges support from the Singapore MOE (Grant No. MOE2018-T3-1-005). L.S. acknowledges support from the Singapore MOE (Grant No. R-265-000-691-114 and MOE2019-T2-2-030, respectively). K. P. L. acknowledges funding from A*STAR AME-IRG program (Grant No: A1983c0035). Y.W., I.A., I.V., G.E., K.P.L. and S.Y.Q. acknowledge support from the Singapore National Research Foundation, Prime Minister's Office, under its medium-sized centre program. Y.W. acknowledges technical help from Miguel Dias Costa and Jun Zhou. Computations were performed on the NUS CA2DM computational cluster and National Supercomputing Centre Singapore. The authors would like to thank Yuan Ping Feng for his comments on the manuscript.

## Author Contributions

S.Y.Q. and L.S. conceived the project. S.Y.Q. led the collaborative effort and direction of the project, and provided guidance on the theoretical analysis and calculations. L.S. provided guidance on the high-throughput piezoelectric calculations. Y.W. performed the calculations and theoretical analysis. I.A. designed the experiments and performed micromechanical cleavage, device fabrication and material characterization under the supervision of K.P.L.. K.C.K, L.W. and I.A. performed the PFM measurements. I.V. synthesized the $NbOX_2$ bulk crystals under the supervision of G.E.. W.H.L performed the LSV measurements under the guidance of K.Y., and both analysed the results. Y.W., I.A. and S.Y.Q. wrote the manuscript. All authors read and commented on the manuscript.

## Competing Interests

The authors declare no competing interests.

## Data Availability Statement

The datasets generated during and/or analysed during the current study are available from the corresponding author on reasonable request.

SUPPLEMENTARY INFORMATION

# Data-driven discovery of high performance layered van der Waals piezoelectric $NbOI_2$

Yaze Wu[1,2,9], Ibrahim Abdelwahab[2,5,9], Ki Chang Kwon[5], Ivan Verzhbitskiy[1,2], Lin Wang[5], Weng Heng Liew[6], Kui Yao[6], Goki Eda[1,2], Kian Ping Loh[2,5,7*], Lei Shen[3,4*], Su Ying Quek[1,2,7,8*]

[1]Department of Physics, National University of Singapore, Singapore, Singapore.

[2]Center for Advanced 2D Materials and Graphene Research Centre, Singapore, Singapore.

[3]Department of Mechanical Engineering, National University of Singapore, Singapore, Singapore.

[4]Engineering Science Programme, National University of Singapore, Singapore, Singapore.

[5]Department of Chemistry, National University of Singapore, Singapore, Singapore.

[6]Institute of Materials Research and Engineering, A*STAR (Agency for Science, Technology and Research), Singapore, Singapore.

[7]NUS Graduate School, Integrative Sciences and Engineering Programme, National University of Singapore, Singapore.

[8]Department of Materials Science and Engineering, National University of Singapore, Singapore.

[9]These authors contributed equally: Yaze Wu, Ibrahim Abdelwahab.

*e-mail: chmlohkp@nus.edu.sg; shenlei@nus.edu.sg; phyqsy@nus.edu.sg

# Table of Contents

# I. Supplementary Tables and Figures

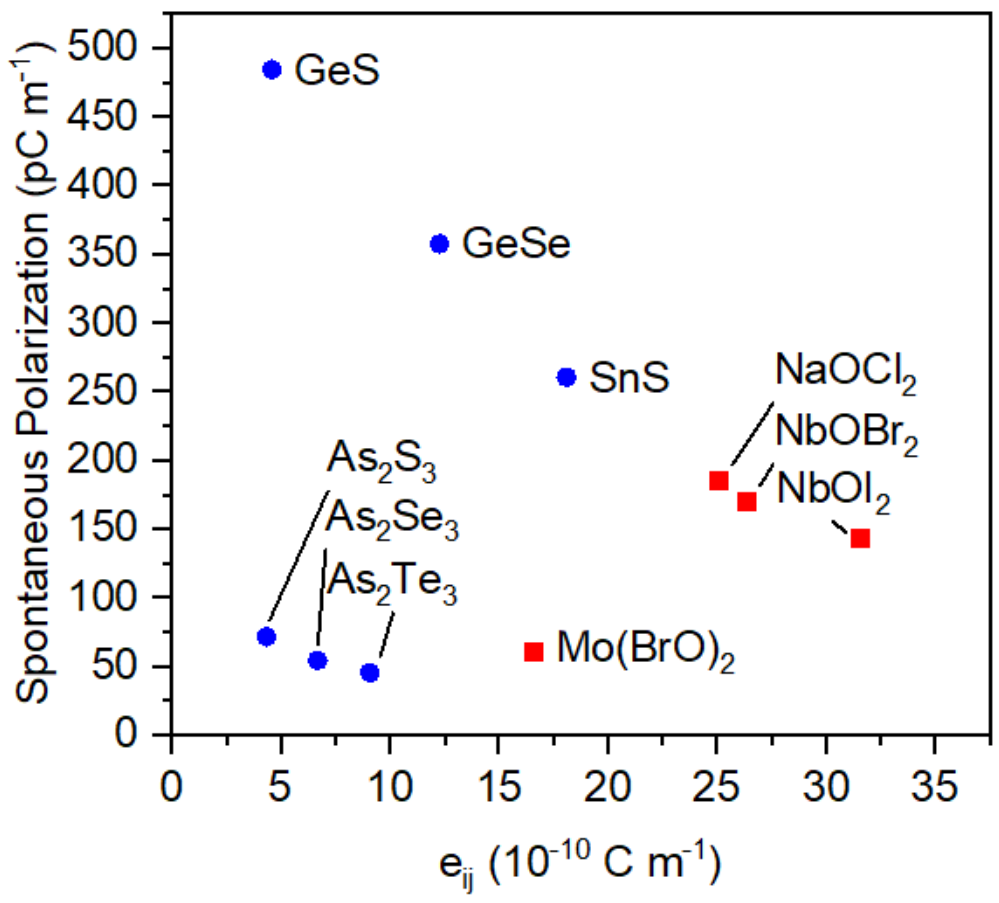

**Fig. S1 | Plot of spontaneous polarization ($pC\ m^{-1}$) and maximum sheet $e_{ij}$ for various 2D materials.** Red squares denote data obtained from this study, blue dots denote data obtained from references[1-3].

**Table S1 | Values of relaxed-ion $e_{ij}$ ($10^{-10}\ C\ m^{-1}$) under different definitions.** RI1 and RI2 are different definitions of relaxed-ion schemes. In RI1, the $e_{ij}$ is obtained using the Berry Phase method by straining one axis, allowing the ionic positions to relax, while not allowing the unstrained lattice vectors to relax. In RI2, the unstrained lattice is allowed to relax to account for the Poisson effect. The values of $e_{ij}$ obtained in these schemes are compared to those obtained through DFPT and by Fei et al. We note that the Poisson effect reduces the effective $e_{22}$ for GeSe significantly.

| Material | DFPT | RI1 | RI2 | Ref |
|---|---|---|---|---|
| GeSe | 12.4 | 12.1 | ~0.3 | 12.3[3] |
| SnSe | 28.2 | 28.6 | 14.1 | 34.9[3] |
| $NbOCl_2$ | 25.4 | 25.8 | 23.3 | |
| $NbOBr_2$ | 26.4 | 27.3 | 27.1 | |
| $NbOI_2$ | 31.6 | 33.3 | 31.4 | |

**Table S2 | Relaxed-ion piezoelectric tensor element ($e_{ij}$), piezoelectric strain tensor element ($d_{ij}$), and spontaneous polarization ($P_i$) of selected materials in units of $10^{-10}\ C\,m^{-1}$, $pm\ V^{-1}$ and $pC\ m^{-1}$ respectively.** In the cases where lattice orientations used in this study is different from existing literature, indices of tensor elements from this study are used.

| | | $e_{ij}$ | | $d_{ij}$ | | $P_i$ |
|---|---|---|---|---|---|---|
| Material | ij | This study | Other studies | This Study | Other studies | |
| 1H-$MoS_2$ | 11 | 3.72 | 3.64[4] | | 3.73[4] | |
| 1H-$MoSe_2$ | 11 | 3.84 | 3.92[4] | | 4.72[4] | |
| 1H-$WS_2$ | 11 | 2.54 | 2.47[4] | | 2.19[4] | |
| 1H-$WSe_2$ | 11 | 2.60 | 2.71[4] | | 2.79[4] | |
| 1H-BN | 11 | 1.46 | 1.38[4] | | 0.60[4] | |
| InSe | 22 | 0.85 | 0.57[5] | | 1.46[5] | |
| GaS | 22 | 1.87 | 1.34[5] | | 2.06[5] | |
| GaSe | 22 | 1.79 | 1.47[5] | | 2.30[5] | |
| $NbOCl_2$ | 11 | 25.1 | | 27.4 | | 185 |
| $NbOBr_2$ | 11 | 26.4 | | 30.0 | | 170 |
| $NbOI_2$ | 11 | 31.6 | | 42.2 | | 143 |
| GeS | 11 | | 4.6[3] | | 75.43[3] | 484[2] |
| GeSe | 11 | 12.3 | 12.3[3] | | 212[3] | 357[2] |
| SnS | 11 | | 18.1[3] | | 144.76[3] | 260[2] |
| $As_2S_3$ | 22 | 1.72 | 4.36[1] | | 55.7[1] | 71[1] |
| $As_2Se_3$ | 22 | 3.15 | 6.71[1] | | 61.7[1] | 54[1] |
| $As_2Te_3$ | | | 9.09[1] | | 61.9[1] | 21.6[1] |
| Bulk CIPS (0K DFT) | 33 | | 1.85[6] | | 18[6] | |
| Bulk CIPS (Exp) | 33 | | 36.7[6] | | 110[6] | |

Here we note the disparity in the $e_{22}$ values for $As_2X_3$ (X=S, Se) obtained in our study and those by Gao et al.[1] and attribute the disparity to the use of different exchange-correlation functionals.

In the figures below, we verify the dynamical stability of $NbOI_2$ by making sure there are no imaginary phonon frequencies for all **q** in the phonon dispersion spectra as well as by ensuring that there is no bond breaking in the molecular dynamics calculation.

The phonon dispersion of $NbOI_2$ is calculated using Quantum Espresso[7-9] working with a Projector Augmented Wave (PAW) approach[10] with the Perdew, Becke and Ernzerhof (PBE) Generalized Gradient Approximation (GGA) of the exchange-correlation functional[11]. We use a kinetic energy cut-off of 60 Ry for the plane wave basis set and a Monkhorst-Pack k point meshes of 12×6×1 is used. The atomic coordinates are fully relaxed using the conjugate gradient scheme until the maximum energy difference between iterations is less than $10^{-14}$ Ry and the residual force is less than 0.0001 Ry/Bohr.

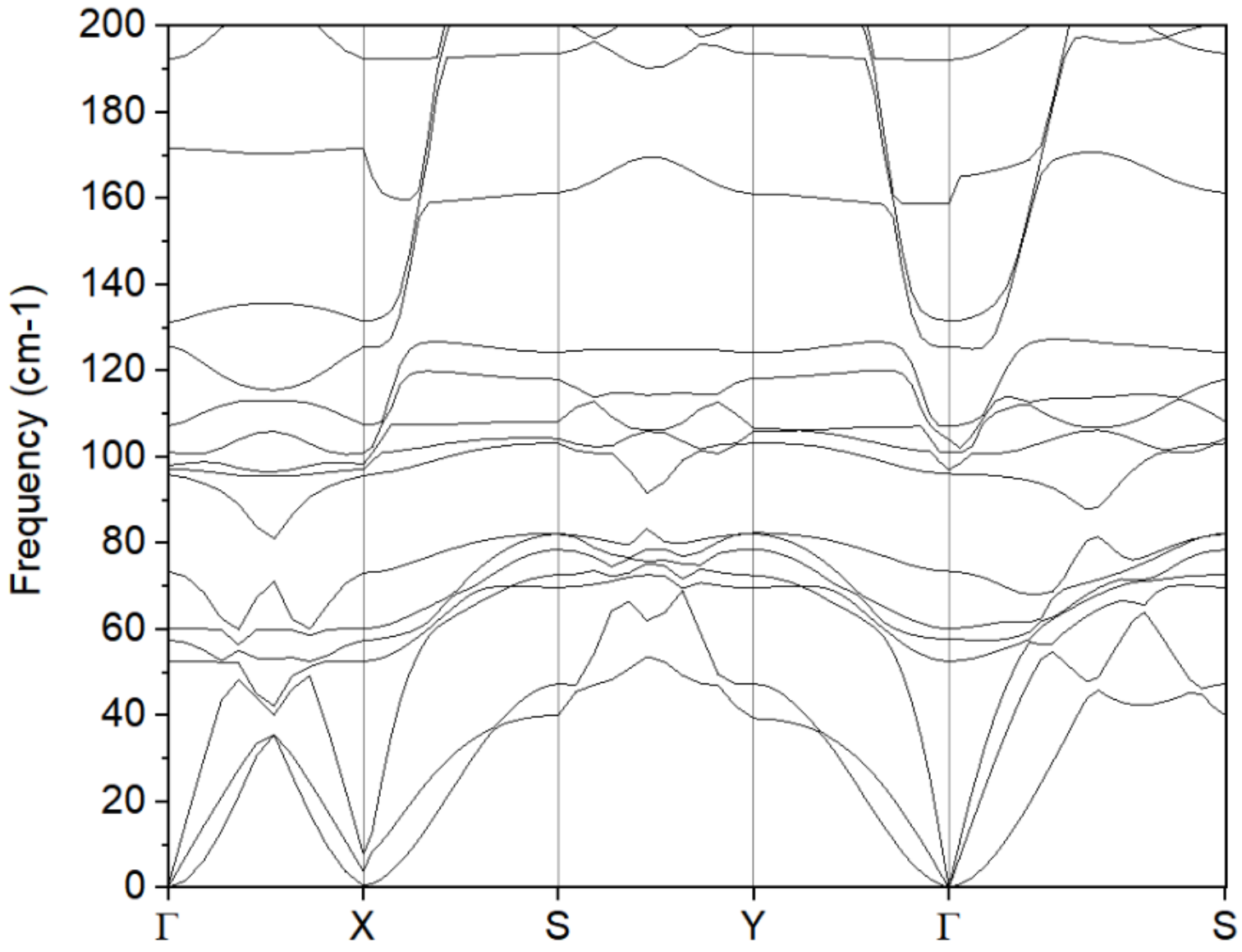

**Fig. S2 | Phonon dispersion of monolayer $NbOI_2$.** No imaginary frequency is observed, hence confirming that the structure is dynamically stable.

Ab initio molecular dynamics of $NbOI_2$ is performed using the VASP code[12] using the same set up as the targeted study on $NbOX_2$. Here, we use a 4×2×1 supercell and a Monkhorst-Pack k point mesh[13] with a density of 3×3×1. We use the NVT ensemble at 298 K with the Nose-Hoover thermostat for 5ps and a timestep of 1fs.

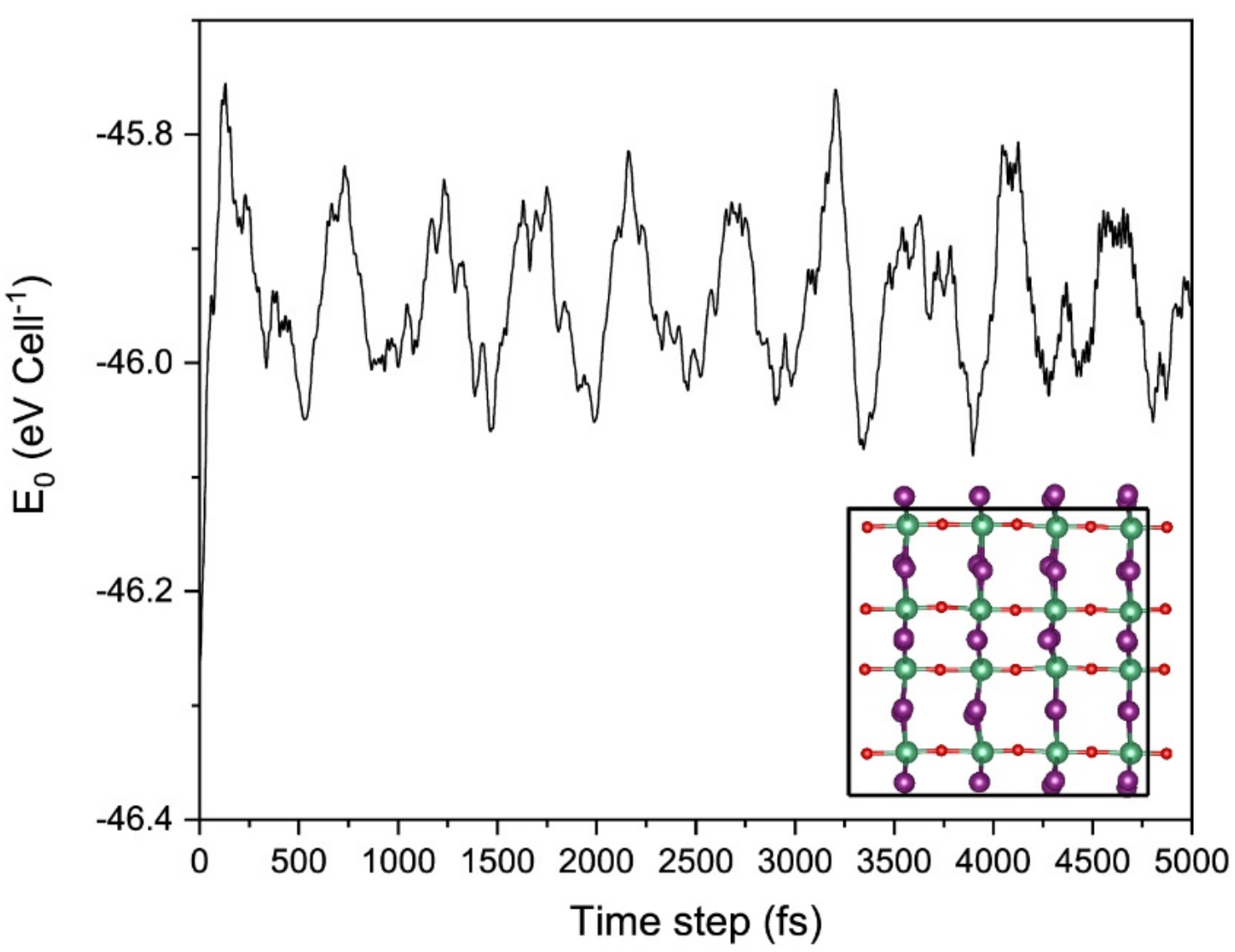

**Fig. S3 | Energy per cell against time step for a molecular dynamics calculation performed on $NbOI_2$ at 298K.** A 4 × 2 supercell is used for the simulation. The total energy of the system fluctuates about a fixed value, suggesting a dynamically stable structure. The inset shows the final structure after 5000 time steps of 1fs each, revealing no broken bonds, hence suggesting a dynamically stable structure at 298K.

**Table S3 | Details of structural parameters for DFT-optimized structures of monolayer $NbOX_2$.** ***a*** and ***b*** denote the in-plane lattice vectors in the polar (*x)* and non-polar (*y)* directions respectively; $l_1$ and $l_2$ are, respectively, the long and short Nb-O bond lengths in Figure 3. The magnitude of $\delta_x^{eqm}$ decreases down the halogen group with $NbOCl_2$ exhibiting the largest Nb displacement from the high symmetry point where $\delta x = 0$ Å and $NbOI_2$ exhibiting the least.

| | $NbOCl_2$ | $NbOBr_2$ | $NbOI_2$ |
|---|---|---|---|
| ***a*** (Å) | 3.964 | 3.964 | 3.973 |
| ***b*** (Å) | 6.677 | 7.096 | 7.595 |
| $l_1$ (Å) | 2.142 | 2.133 | 2.123 |
| $l_2$ (Å) | 1.824 | 1.831 | 1.850 |
| $\delta_x^{eqm}$ (Å) | 0.318 | 0.302 | 0.274 |

**Table S4 | Piezoelectric stress tensor elements ($e_{11}$) and piezoelectric strain tensor elements ($d_{11}$) of bulk $NbOX_2$.** $\boldsymbol{e}_{11}$ of the bulk $NbOX_2$ is presented in both bulk units ($C\ m^{-2}$) as well as sheet units ($10^{-10}\ C\ m^{-1}$), where for comparison with $\boldsymbol{e}_{11}$ in the monolayer, we multiply the bulk $\boldsymbol{e}_{11}$ by the cell height to obtain an equivalent sheet value.

| | $\boldsymbol{e}_{11}$ ($C\ m^{-2}$) | sheet $\boldsymbol{e}_{11}$ ($10^{-10}\ C\ m^{-1}$) | $\boldsymbol{d}_{11}$ ($pC\ N^{-1}$) |
|---|---|---|---|
| $NbOCl_2$ | 3.56 | 25.3 | 27.4 |
| $NbOBr_2$ | 3.46 | 26.6 | 30.1 |
| $NbOI_2$ | 3.76 | 31.3 | 42.0 |

**Table S5 | Piezoelectric tensor elements and electromechanical coupling factor of bulk $NbOX_2$.** Here we report values obtained from two approximations (PBE and PBE-D3) to the exchange-correlation functional.

| | PBE | | | PBE-D3 | | |
|---|---|---|---|---|---|---|
| | $NbOCl_2$ | $NbOBr_2$ | $NbOI_2$ | $NbOCl_2$ | $NbOBr_2$ | $NbOI_2$ |
| $\boldsymbol{e}_{11}$ $(C\,m^{-2})$ | 3.6 | 3.4 | 3.8 | 4.5 | 4.6 | 5.3 |
| $\boldsymbol{e}_{12}$ $(C\,m^{-2})$ | -0.1 | -0.1 | -0.1 | -0.2 | -0.2 | -0.2 |
| $\boldsymbol{e}_{13}$ $(C\,m^{-2})$ | -0.1 | 0.0 | 0.0 | -0.1 | -0.1 | -0.1 |
| $\boldsymbol{e}_{26}$ $(C\,m^{-2})$ | 0.1 | 0.1 | 0.1 | 0.1 | 0.1 | 0.1 |
| $\boldsymbol{e}_{35}$ $(C\,m^{-2})$ | 0.0 | 0.0 | 0.0 | 0.0 | 0.0 | 0.0 |
| $\boldsymbol{d}_{11}$ $(pm\,V^{-1})$ | 27.3 | 30.1 | 41.8 | 27.3 | 30.2 | 43.6 |
| $\boldsymbol{d}_{12}$ $(pm\,V^{-1})$ | -4.0 | -3.7 | -4.7 | -4.2 | -4.2 | -5.5 |
| $\boldsymbol{d}_{13}$ $(pm\,V^{-1})$ | -16.9 | -21.0 | -21.7 | -11.0 | -10.6 | -13.2 |
| $\boldsymbol{d}_{26}$ $(pm\,V^{-1})$ | 5.9 | 6.1 | 5.4 | 5.6 | 5.9 | 5.0 |
| $\boldsymbol{d}_{35}$ $(pm\,V^{-1})$ | 4.1 | 4.9 | 4.0 | 1.2 | 1.5 | 1.5 |
| $k$ | 0.95 | 0.96 | 1.07 | 0.93 | 0.92 | 1.02 |

**Table S6 | Dielectric constants of monolayer and bulk $NbOX_2$, computed by DFT.** Values for monolayer are caluclated with PBE and is scaled with respect to the thickness of the monolayer while values for bulk are calculated with both PBE and PBE-D3.

| | Monolayer (PBE) | | | Bulk (PBE) | | | Bulk (PBE-D3) | | |
|---|---|---|---|---|---|---|---|---|---|
| | $NbOCl_2$ | $NbOBr_2$ | $NbOI_2$ | $NbOCl_2$ | $NbOBr_2$ | $NbOI_2$ | $NbOCl_2$ | $NbOBr_2$ | $NbOI_2$ |
| $\boldsymbol{\varepsilon}_{xx}$ | 11.9 | 12.5 | 15.7 | 12.1 | 12.6 | 15.5 | 16.2 | 18.4 | 25.0 |
| $\boldsymbol{\varepsilon}_{yy}$ | 11.1 | 11.1 | 11.9 | 11.2 | 11.1 | 12.0 | 13.5 | 13.8 | 14.4 |
| $\boldsymbol{\varepsilon}_{zz}$ | 1.9 | 1.9 | 2.0 | 3.2 | 3.5 | 4.2 | 4.3 | 5.1 | 6.3 |

**Table S7 | Stiffness Tensor elements $C_{11}$, $C_{12}$, $C_{22}$ of monolayer $NbOX_2$ in the unit of N $m^{-1}$ and Compliance tensor elements $S_{11}$, $S_{12}$, $S_{22}$, $S_{66}$ in m $N^{-1}$.**

| | $C_{11}$ | $C_{12}$ | $C_{22}$ | $C_{66}$ | $S_{11}$ | $S_{12}$ | $S_{22}$ | $S_{66}$ |
|---|---|---|---|---|---|---|---|---|
| $NbOCl_2$ | 92.9 | 5.3 | 62.0 | 15.2 | 10.8 | -1.1 | 15.5 | 65.9 |
| $NbOBr_2$ | 89.0 | 5.6 | 63.1 | 14.4 | 11.3 | -1.0 | 15.9 | 69.2 |
| $NbOI_2$ | 75.6 | 5.3 | 62.0 | 13.6 | 13.3 | -1.1 | 16.2 | 73.5 |

It can be observed that the Young's Modulus along the $a_x$ axis (i.e. $C_{11}$) decreases by about 19% down the halogen group from 92.9 N $m^{-1}$ for $NbOCl_2$ to 75.6 N $m^{-1}$ for $NbOI_2$, while that along $a_y$ axis ($C_{22}$) does not change significantly.

**Table S8 | Summary of piezoelectric coefficients from experiments.** The precision of the magnitudes are kept at the levels reported in the references.

| Material | Piezoelectric coefficient | Magnitude | Reference |
|---|---|---|---|
| h-$MoS_2$ | $e_{11}$ | $2.9 \times 10^{-10}$ C $m^{-1}$ | 14 |
| h-BN | $e_{11}$ | $2.91 \times 10^{-10}$ C $m^{-1}$ | 15 |
| h-MoSSe | $d_{33}$ | 0.1 pm $V^{-1}$ | 16 |
| $WTe_2$ | $d_{33}$ | 0.7 pm $V^{-1}$ | 17 |
| g-$C_3N_4$ | $d_{33}$ | 1 pm $V^{-1}$ | 18 |
| CdS | $d_{33}$ | 16.4 pm $V^{-1}$ | 19 |
| SnSe | $d_{11}$ | 23 pm $V^{-1}$ | 20 |
| ZnO nanobelt | $d_{33}$ | 14.3 – 26.7 pm $V^{-1}$ | 21 |
| $\alpha$-$In_2Se_3$ | $d_{33}$ | 5.3 pm $V^{-1}$ | This Work |
| $CuInP_2S_6$ | $d_{33}$ | 4.1 pm $V^{-1}$ | This Work |
| $NbOCl_2$ | $d_{11}$ | 9.4 pm $V^{-1}$ | This Work |
| $NbOI_2$ | $d_{11}$ | 21.45 pm $V^{-1}$ | This Work |

**Table S9 | Comparison between 2D and 3D piezoelectric strain moduli $|d_{ij}|_{max}$ of 2D materials and their corresponding bulk piezoelectric parents.** Values for 2D $|d_{ij}|_{max}$ are obtained as from **Fig. 2b**. Values of 3D $|d_{ij}|_{max}$ for InSe and GaSe are obtained from Materials Project; Values of 3D $|d_{ij}|_{max}$ for $NbOX_2$ are obtained from $d_{11_{PBE-D3}}$ of **Table S5**. The materials project material ID of the parent materials are presented in square brackets.

| *Material* | 2D $|d_{ij}|_{max}$ $(pm\ V^{-1})$ | 3D $|d_{ij}|_{max}$ $(pm\ V^{-1})$ |
|---|---|---|
| *InSe* | 1.5 | 2.3 [mp-22691] |
| *GaSe* | 2.3 | 2.9 [mp-11342]<br>3.3 [mp-1572] |
| *$NbOCl_2$* | 27.4 | 27.3 [mp-549720] |
| *$NbOBr_2$* | 30.0 | 30.2 [mp-550070] |
| *$NbOI_2$* | 42.2 | 43.6 [mp-1025567] |

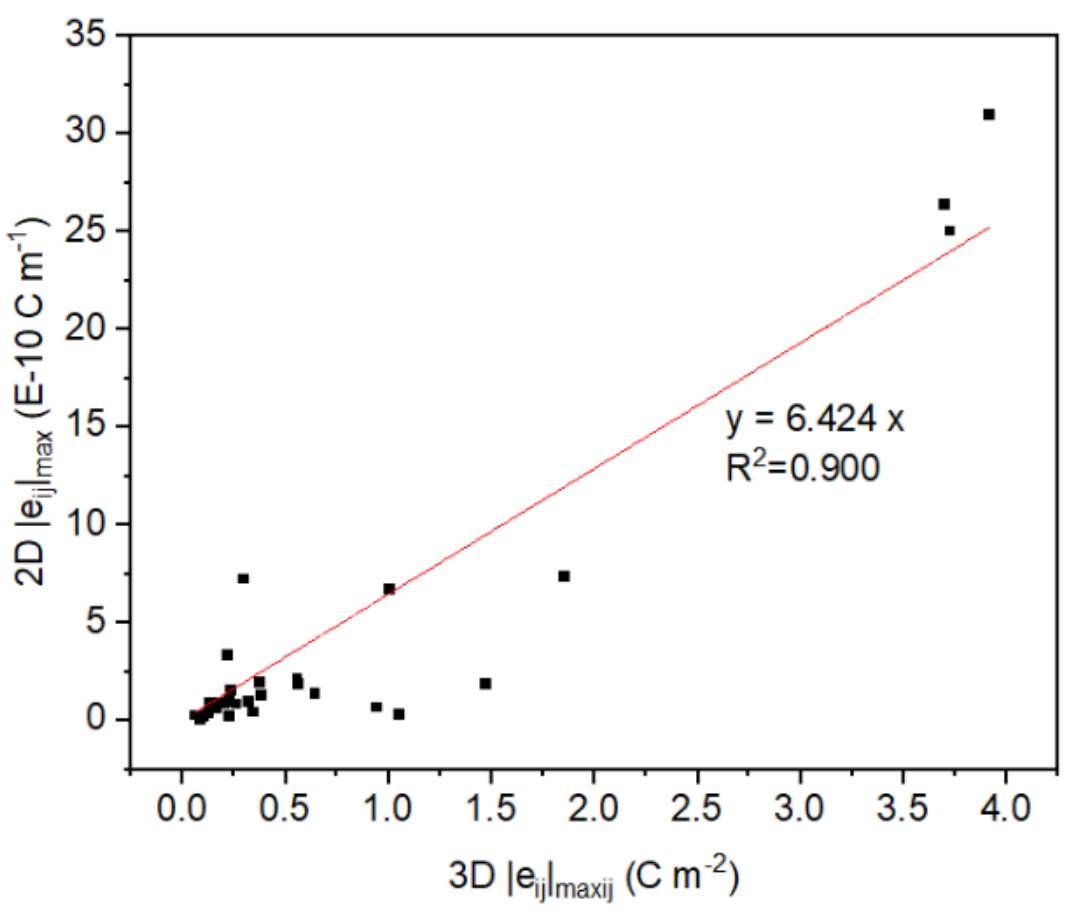

**Fig. S4 | Correlation between 2D and 3D piezoelectric stress moduli $|e_{ij}|_{max}$ of 2D materials and their corresponding bulk piezoelectric parents.** The linear trend line passing through the origin has a gradient of ~6.4Å. Numerical details of each datapoint can be obtained from **Table S10**. Unlike the piezoelectric strain coefficients, the piezoelectric stress coefficients are defined differently for 2D and 3D systems, as described in the Formalisms section. The gradient of ~6.4Å corresponds to the average thickness of a monolayer.

**Table S10 | Numerical values of 2D and 3D piezoelectric stress moduli $|e_{ij}|_{max}$ of 2D materials and their corresponding bulk piezoelectric parents, as presented in Fig. S4.** Headings of the first 3 columns are the exact database dictionary keys used in this database. Heading of the last column is in the format MP_{dictionary_key_in_materials_project}.

| *formula_pretty* | *mat_project_id* | *dielectric.max_abs_sheet_piezo* (E-10 $C\ m^{-1}$) | *MP_piezo.eij_max* ($C\ m^{-2}$) |
|---|---|---|---|
| $Li_2WS_4$ | mp-753195 | 0.0583 | 0.0902 |
| $Cu_2WSe_4$ | mp-1025340 | 0.2071 | 0.2319 |
| $LiBH_4$ | mp-644223 | 0.2221 | 0.1023 |
| $Cu_2WS_4$ | mp-8976 | 0.2314 | 0.0621 |
| $NaHO$ | mp-626000 | 0.2808 | 1.0514 |
| $Sn(PS_3)_2$ | mp-36381 | 0.3645 | 0.1243 |
| $H_3BrO$ | mp-625521 | 0.3860 | 0.3468 |
| $ZrCl_2$ | mp-23162 | 0.6390 | 0.1678 |
| $CaHClO$ | mp-642725 | 0.6500 | 0.9465 |
| $BiTeCl$ | mp-28944 | 0.8225 | 0.2654 |
| $InSe$ | mp-22691 | 0.8475 | 0.1331 |
| $Ta_3TeI_7$ | mp-29117 | 0.8632 | 0.2082 |
| $Nb_3SBr_7$ | mp-29057 | 0.8645 | 0.1626 |
| $Hg_3AsSe_4Br$ | mp-567949 | 0.9225 | 0.3205 |
| $Nb_3TeI_7$ | mp-29689 | 0.9446 | 0.2286 |
| $BiTeBr$ | mp-33723 | 1.2412 | 0.3846 |
| $Hg_2P_2S_7$ | mp-27171 | 1.3647 | 0.6476 |
| $B_2S_2O_9$ | mp-1019509 | 1.5431 | 0.2371 |
| $ZrGeTe_4$ | mp-13542 | 1.8207 | 0.5666 |
| $AlHO_2$ | mp-625054 | 1.8717 | 1.4761 |
| $BiTeI$ | mp-22965 | 1.9007 | 0.3763 |
| $HfGeTe_4$ | mp-567817 | 2.1278 | 0.5587 |
| $InGaS_3$ | mp-19885 | 3.3485 | 0.2260 |
| $NbTlBr_4O$ | mp-551826 | 6.7080 | 1.0059 |
| $Sn_2IF_3$ | mp-27167 | 7.2032 | 0.3021 |
| $InSnCl_3$ | mp-998560 | 7.3495 | 1.8560 |
| $NbCl_2O$ | mp-1025567 | 24.9742 | 3.7281 |
| $NbBr_2O$ | mp-550070 | 26.3286 | 3.7007 |
| $NbI_2O$ | mp-549720 | 30.9221 | 3.9211 |
| $SbF_3$ | mp-1880 | 46.0720 | 1.5282 |

**Table S11 | Details of structural parameters for bulk $NbOX_2$.** DFT values are compared with those deduced from single-crystal X-ray diffraction in this work and in the ICSD database[22, 23]. **a** and ***b*** denote the in-plane lattice vectors in the polar (*x*) and non-polar (*y*) directions respectively, ***c*** denotes the out-of-plane lattice vector; $\alpha$, $\beta$ and $\gamma$ denote the angles ∠**bc,** ∠**ac** and ∠**ab** respectively. $l_1$ and $l_2$ are, respectively, the long and short Nb-O bond lengths. The magnitude of $\delta_x^{eqm}$ decreases down the halogen group with $NbOCl_2$ exhibiting the largest Nb displacement from the high symmetry point where $\delta x = 0$ Å and $NbOI_2$ exhibiting the least.

| | DFT (PBE) | | | DFT (PBE-D3) | | | Experiment | | ICSD | |
|---|---|---|---|---|---|---|---|---|---|---|
| | $NbOCl_2$ | $NbOBr_2$ | $NbOI_2$ | $NbOCl_2$ | $NbOBr_2$ | $NbOI_2$ | $NbOCl_2$ | $NbOI_2$ | $NbOBr_2$ | $NbOI_2$ |
| **a** (Å) | 3.963 | 3.963 | 3.974 | 3.939 | 3.932 | 3.943 | 3.904 | 3.933 | 3.908 | 3.924 |
| **b** (Å) | 6.773 | 7.096 | 7.601 | 6.744 | 7.065 | 7.558 | 6.720 | 7.523 | 7.023 | 7.520 |
| **c** (Å) | 15.180 | 16.397 | 17.669 | 13.568 | 14.553 | 15.914 | 12.863 | 15.188 | 13.833 | 15.184 |
| $\alpha$ (°) | 104.3 | 105.0 | 105.3 | 105.1 | 105.0 | 104.9 | 105.6 | 105.4 | 105.0 | 105.5 |
| $\beta$ (°) | 105.1 | 104.0 | 103.0 | 90.0 | 90.0 | 90.0 | 90.0 | 90.0 | 90.0 | 90.0 |
| $\gamma$ (°) | 90.0 | 90.0 | 90.0 | 90.0 | 90.0 | 90.0 | 90.0 | 90.0 | 90.0 | 90.0 |
| $l_1$ (Å) | 2.140 | 2.132 | 2.125 | 2.111 | 2.094 | 2.081 | 2.125 | 2.106 | 2.110 | 2.110 |
| $l_2$ (Å) | 1.824 | 1.831 | 1.850 | 1.829 | 1.840 | 1.863 | 1.779 | 1.827 | 1.800 | 1.810 |
| $\delta_x^{eqm}$ (Å) | 0.316 | 0.301 | 0.276 | 0.282 | 0.254 | 0.218 | 0.346 | 0.279 | 0.310 | 0.300 |

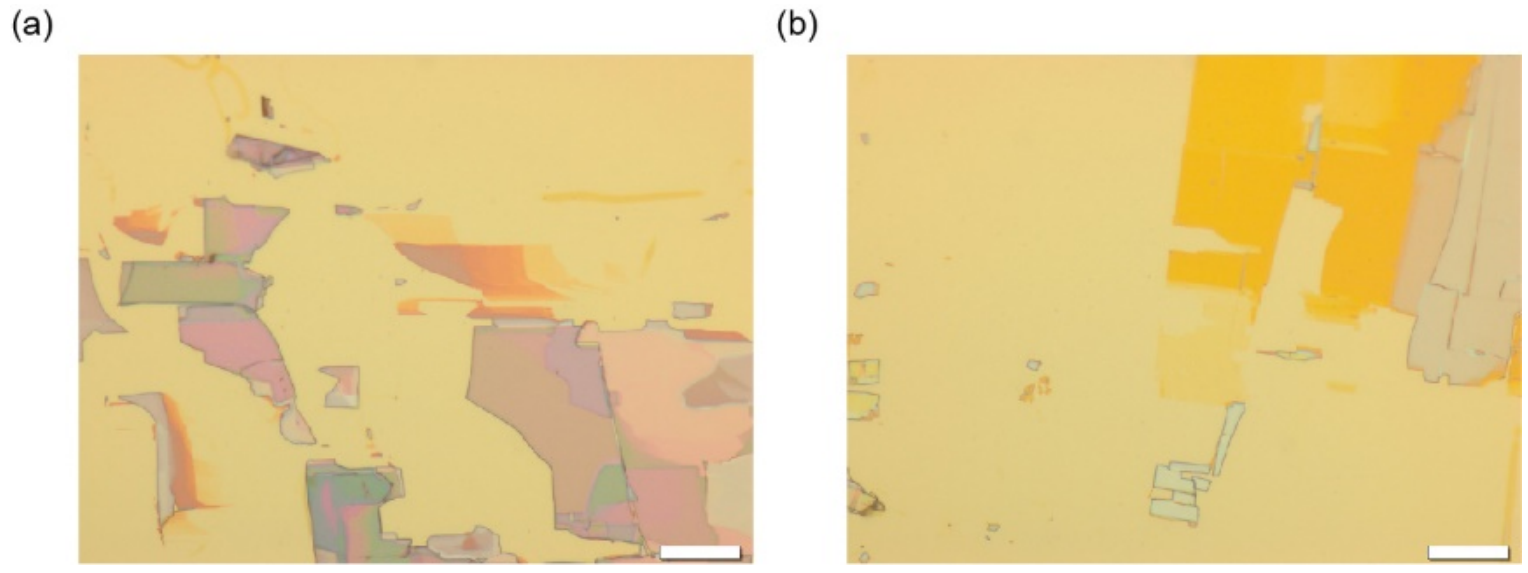

**Fig. S5 | Optical images of exfoliated $NbOX_2$ nanosheets**. (a) $NbOI_2$ flakes on Au substrate. (b) $NbOCl_2$ flakes on Au substrate.

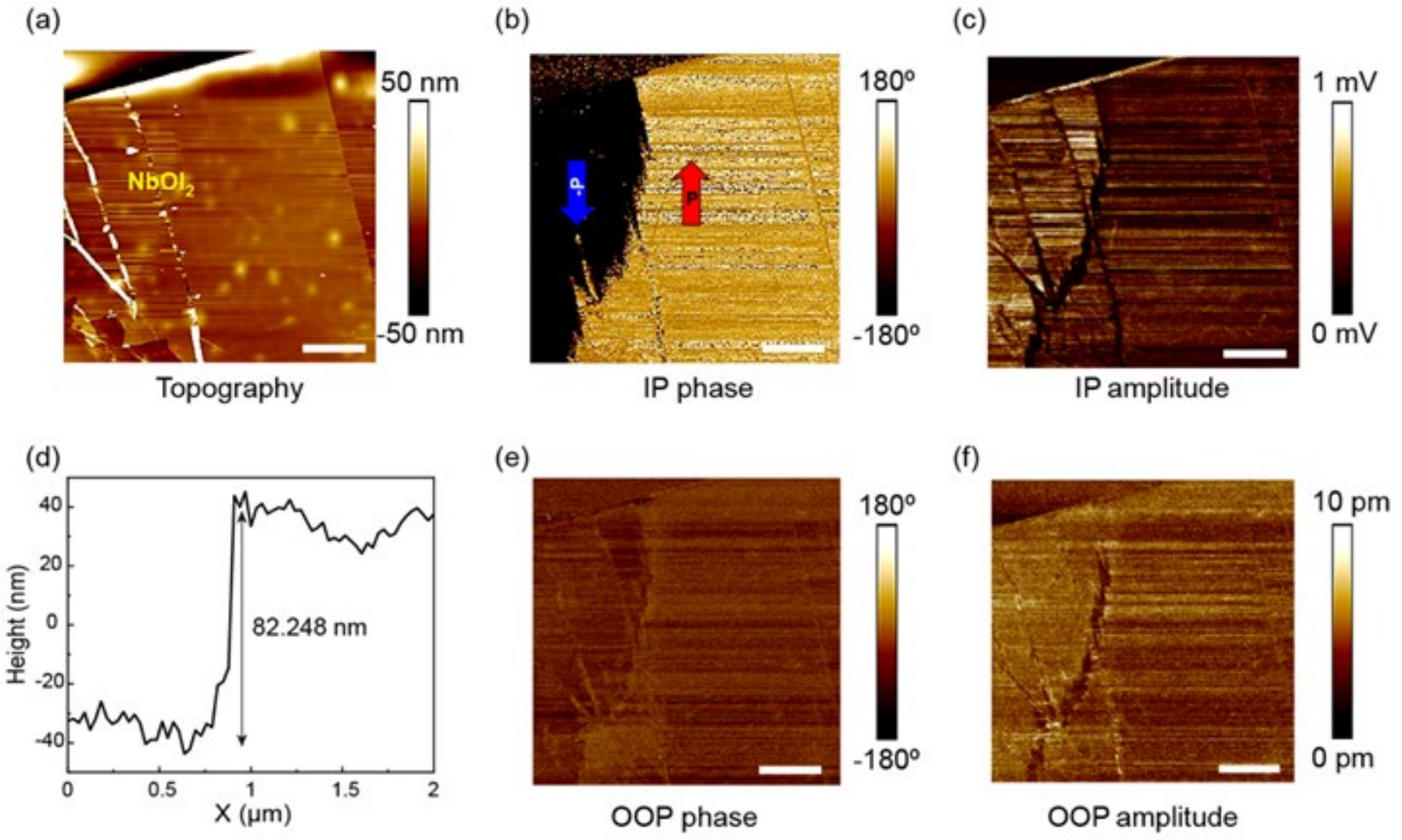

**Fig. S6 | PFM measurements on 82-nm-thick $NbOI_2$.** Topography, height profile, in-plane (IP) phase, IP amplitude, out-of-plane (OOP) phase, and OOP amplitude of 82-nm-thick $NbOI_2$ flake. Scale bars: 4 μm. Drive voltage: 5 V. Drive frequency: 65 kHz.

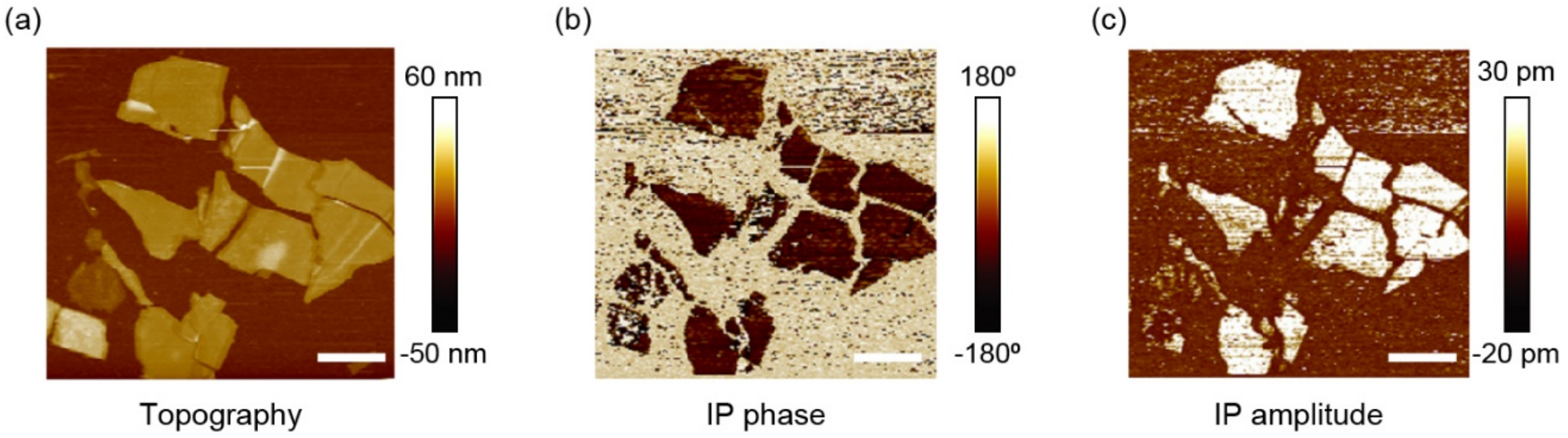

**Fig. S7 | PFM measurements on 23-nm-thick $NbOI_2$.** Topography, in-plane (IP) phase, and IP amplitude images of 23-nm-thick $NbOI_2$ flake. Scale bars: 4 μm. Drive voltage: 2.5 V. Drive frequency: 866 kHz.

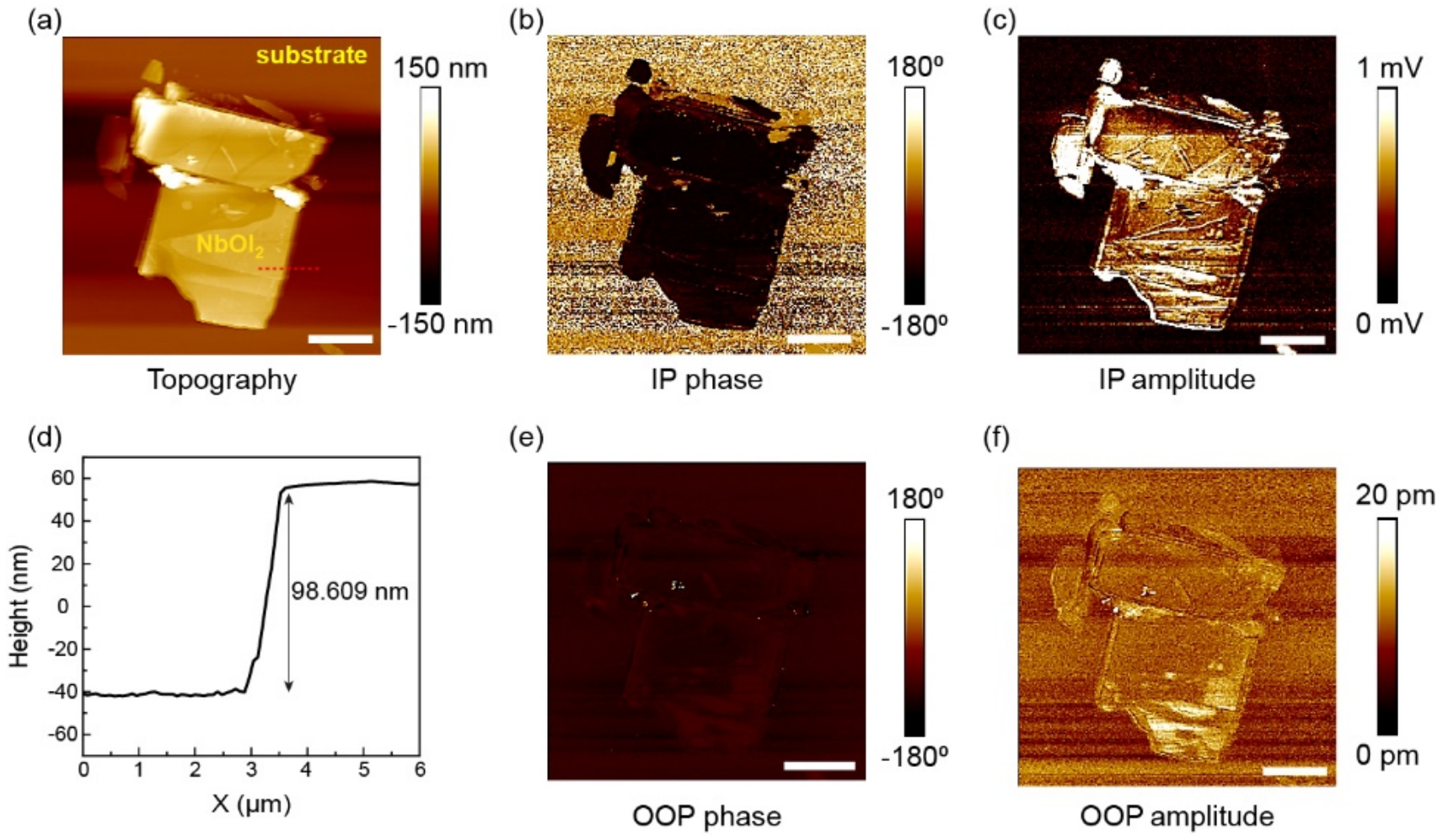

**Fig. S8 | PFM measurements on 98-nm-thick $NbOI_2$.** Topography, height profile, in-plane (IP) phase, IP amplitude, out-of-plane (OOP) phase, and OOP amplitude of 98-nm-thick $NbOI_2$ flake. Scale bars: 4 μm. Drive voltage: 10 V. Drive frequency: 75 kHz.

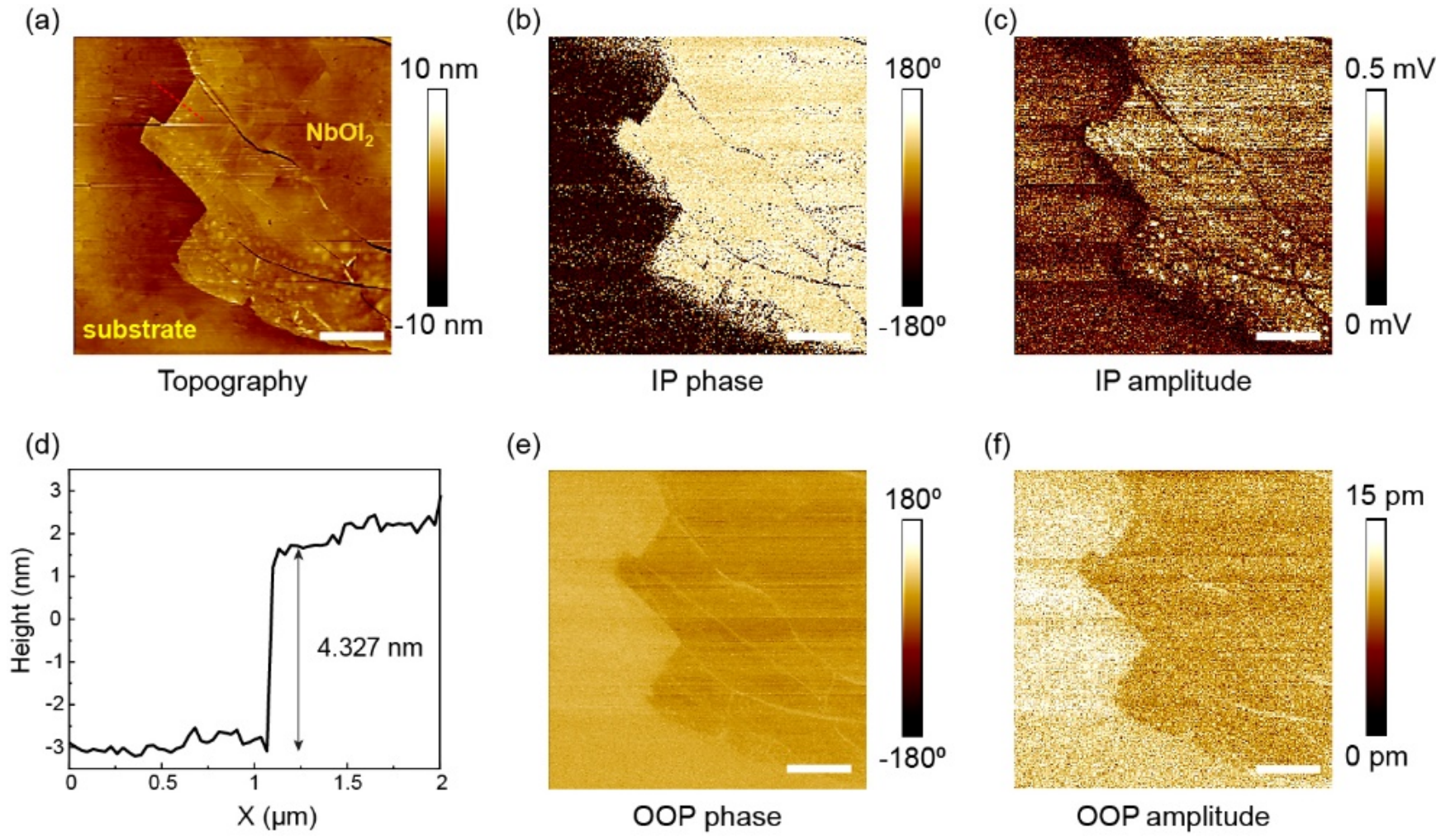

**Fig. S9 | PFM measurements on 4.3-nm-thick $NbOI_2$.** Topography, height profile, in-plane (IP) phase, IP amplitude, out-of-plane (OOP) phase, and OOP amplitude of 4.3-nm-thick $NbOI_2$ flake. Scale bars: 2 μm. Drive voltage: 10 V. Drive frequency: 30.5 kHz.

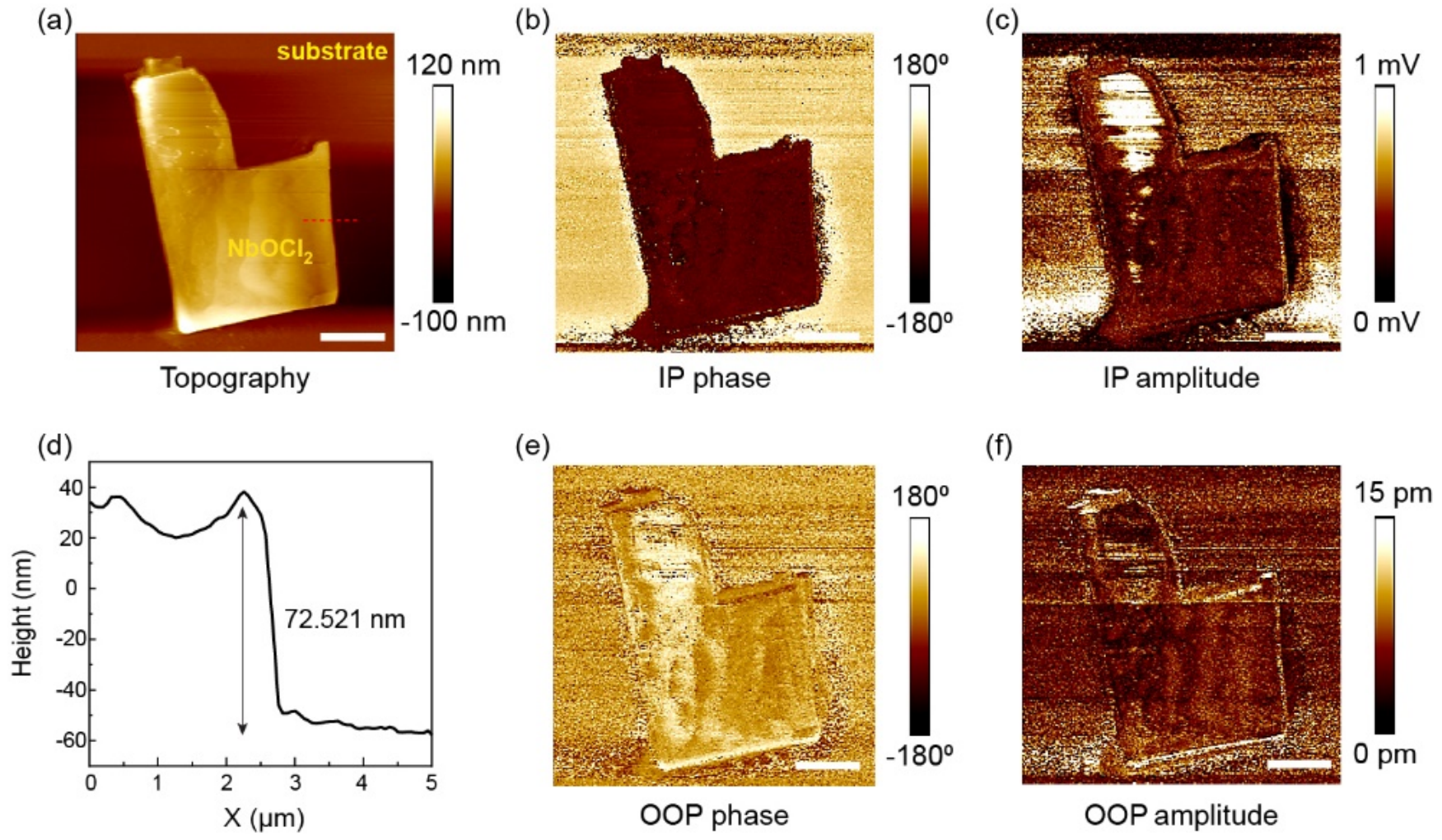

**Fig. S10 | PFM measurements on 72-nm-thick $NbOCl_2$.** Topography, height profile, in-plane (IP) phase, IP amplitude, out-of-plane (OOP) phase, and OOP amplitude of 72-nm-thick $NbOCl_2$ flake. Scale bars: 3 μm. Drive voltage: 10 V. Drive frequency: 75 kHz.

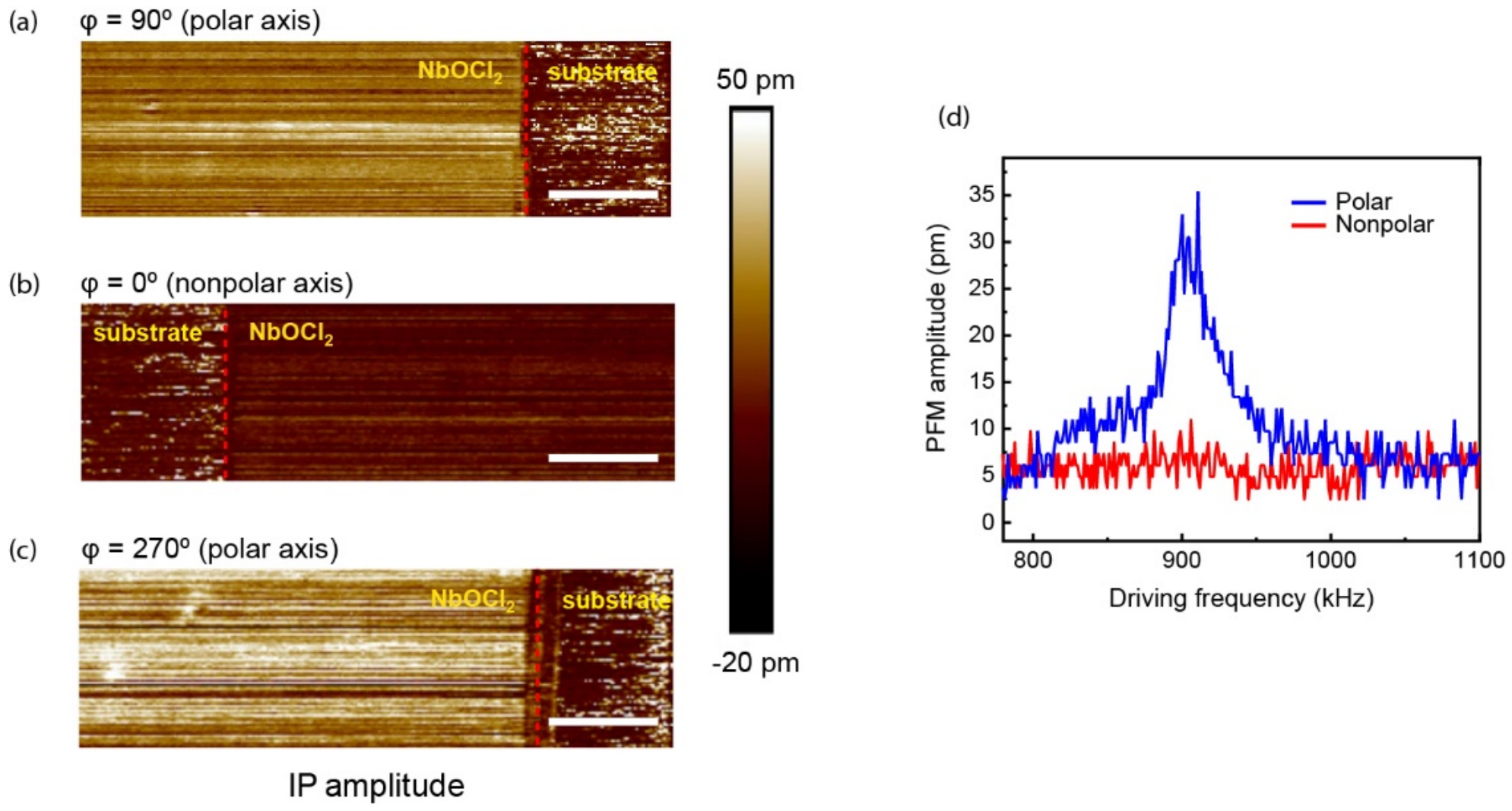

**Fig. S11 | Vector PFM Measurements on 17.2-nm-thick $NbOCl_2$. (a-c)** Vector PFM IP amplitude images of 17.2-nm-thick $NbOCl_2$ showing spontaneous polarization at 90° (a), 0° (b), and 270° (c) angles relative to the cantilever long axis. (**d**) PFM amplitude profiles along the polar and nonpolar axes of the $NbOCl_2$ flake. Scale bars are 2 μm.

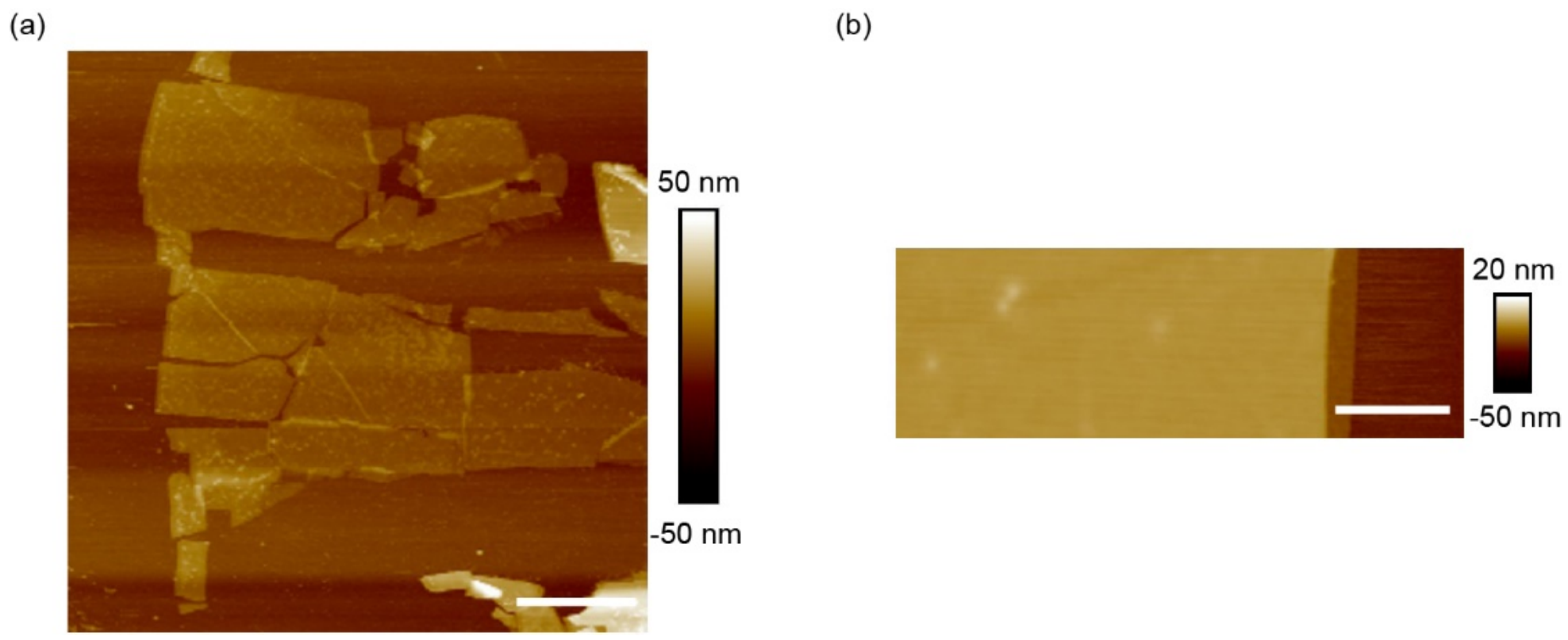

**Fig. S12 | AFM measurements on $NbOX_2$.** Topography images of (a) the 10-nm-thick $NbOI_2$ flakes shown in **Fig. 4**e-g and (b) the 17.2-nm-thick $NbOCl_2$ flake shown in **Fig. S11**. Scale bars: (a) 4 μm, (b) 2 μm.

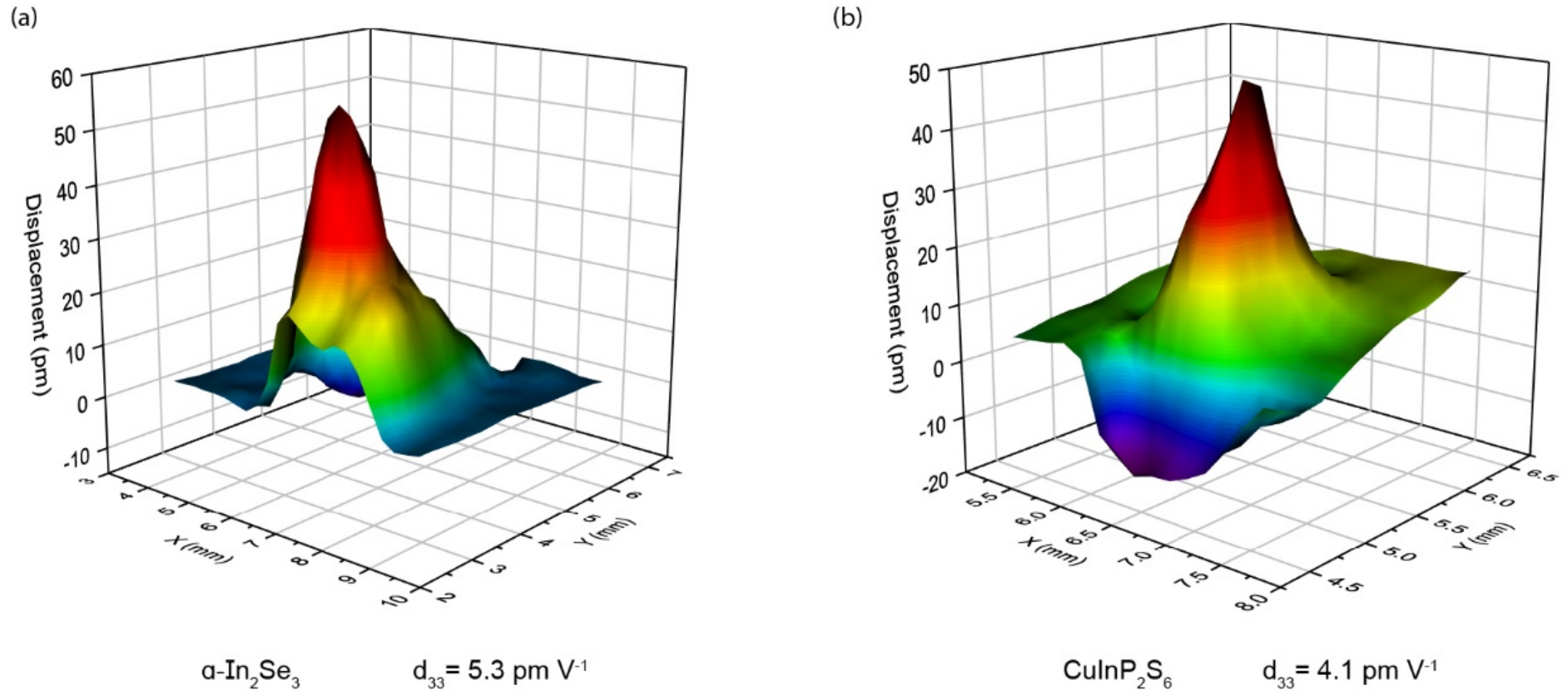

**Fig. S13 | Measurement of the piezoelectric coefficients of α-$In_2Se_3$ and $CuInP_2S_6$ (CIPS) using a laser scanning vibrometer (LSV).** 3D graphs of the instantaneous vibration when the displacement magnitude reaches the maximum under the sine-wave driving electrical signal. The measurements are conducted along the vertical polar directions $d_{33}$ of α-$In_2Se_3$ (a) and CIPS (b). The in-plane cofficients of α-$In_2Se_3$ are also measured and found to be $\boldsymbol{d}_{11}$ = 1.7 pm/V and $\boldsymbol{d}_{22}$ = 6.0 pm/V.

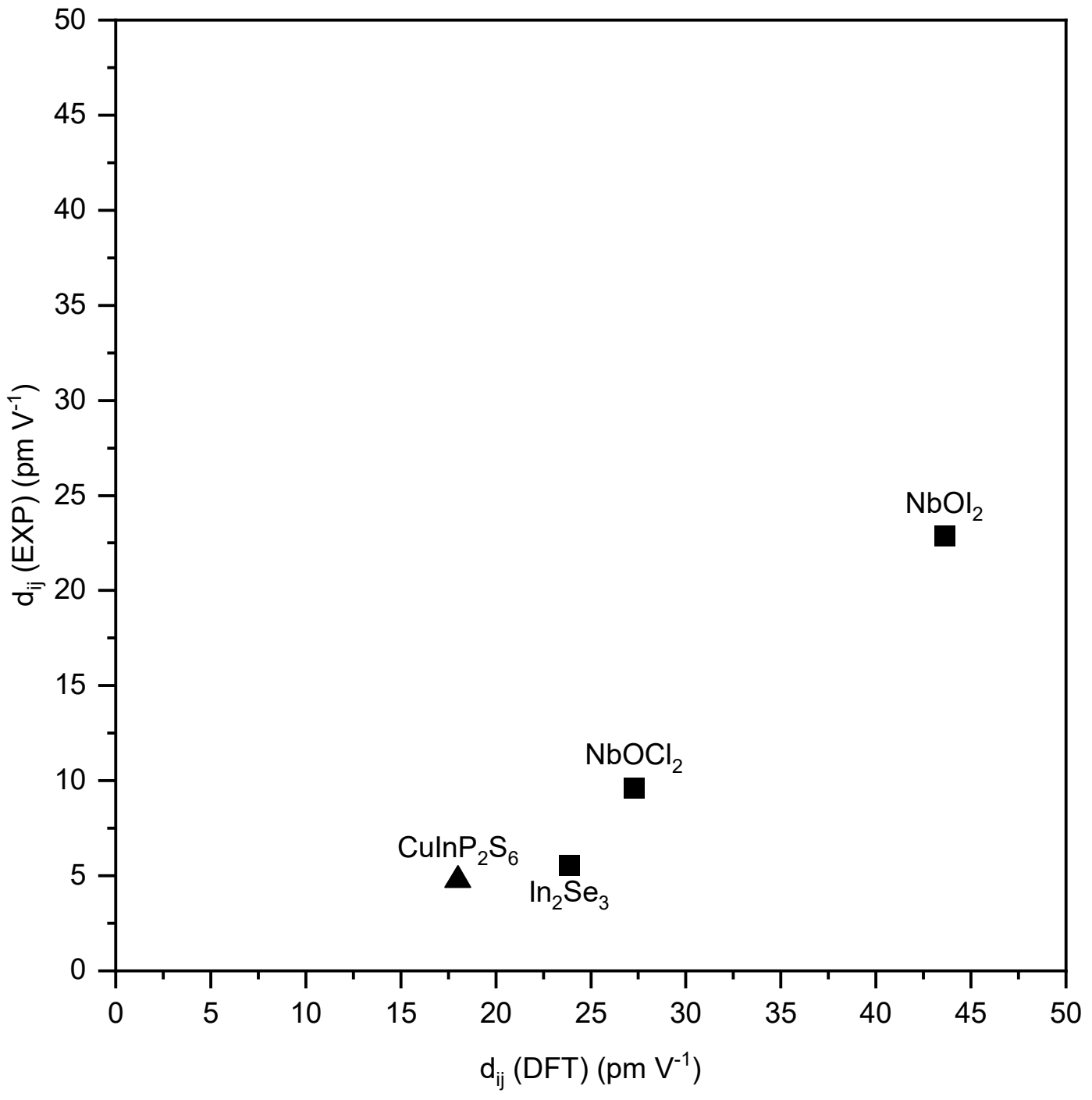

**Fig. S14 | Comparison between theoretical (DFT) and experimental maximal piezoelectric strain tensor elements ($d_{ij}$).** DFT value for $d_{33}$ of $CuInP_2S_6$ is taken from ref [6]. Values for the other materials are obtained in this work. The horizontal and vertical axes of this figure are plotted in scale.

**Table S12 | Electronic and ionic contributions to $e_{11}$ of $NbOX_2$.** The electronic contribution (e) refers to value of $e_{11}$ computed from the "clamped ion" configuration.

| $e_{11}$ | e | ion | total |
|---|---|---|---|
| $NbOCl_2$ | -0.4 | 25.5 | 25.1 |
| $NbOBr_2$ | -0.4 | 26.8 | 26.4 |
| $NbOI_2$ | -0.2 | 31.8 | 31.6 |

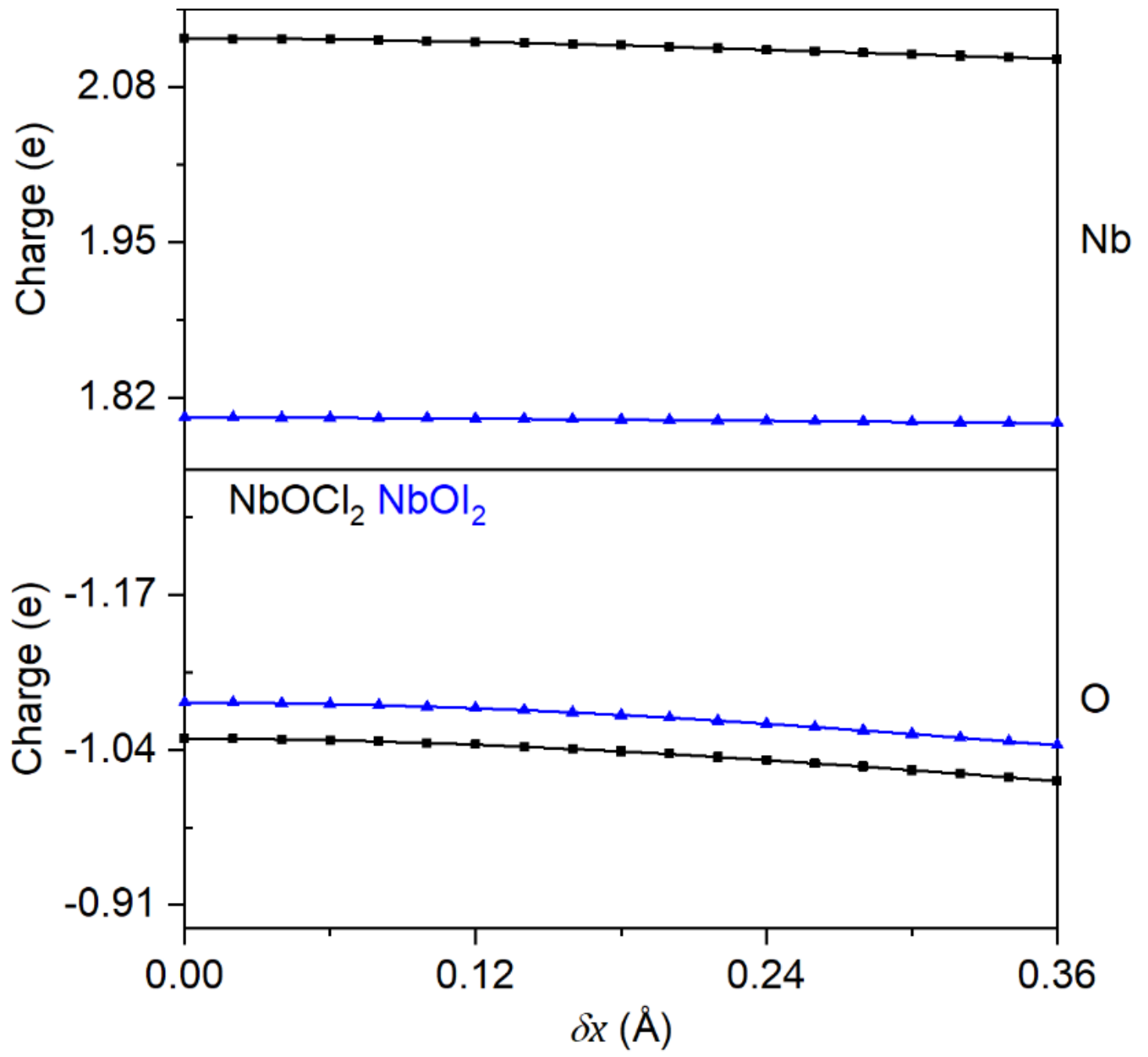

**Fig. S15 | Static charges on Nb and O atoms in $NbOI_2$ and $NbOCl_2$, plotted as a function of $\boldsymbol{\delta x}$.** These charges are computed used the Bader approach[24-27].

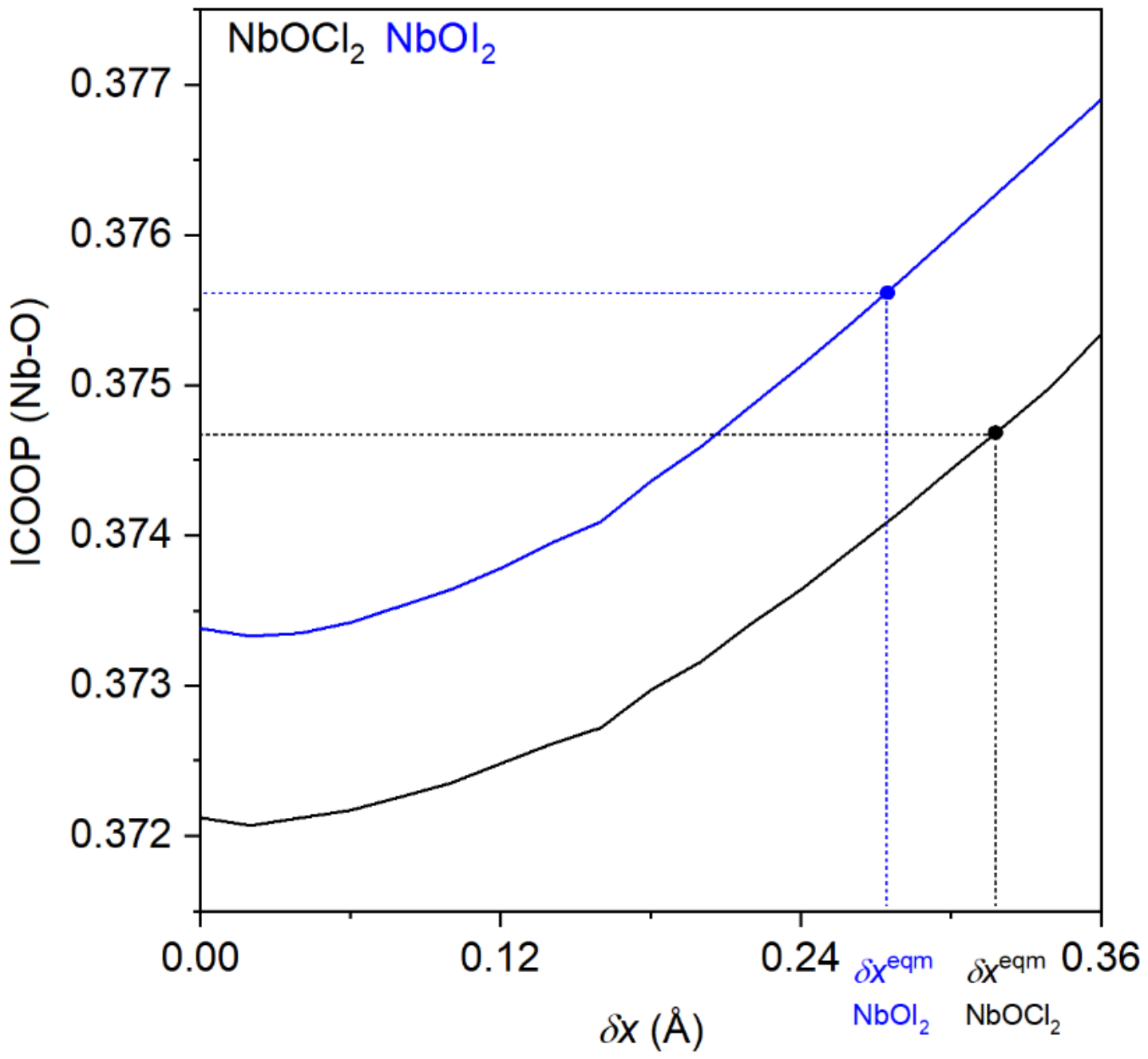

**Fig. S16 | Integrated crystal orbital overlap population (ICOOP) for the Nb-O bond in $NbOX_2$.** ICOOP is a measure of the degree of covalency in a bond, and a more positive value indicates greater covalency[28]. Increasing covalence in the Nb-O bond is observed as Nb is moved from the centred symmetric structure, with increasing $\delta x$. This observation is consistent with the pseudo-Jahn-Teller effect. At the same $\delta x$, Nb-O bonds are more covalent in $NbOI_2$ than in $NbOCl_2$, which can be explained by the larger electronegativity of Cl compared to I. Dotted lines denote the values for the equilibrium structures. Nb-O bonds in $NbOI_2$ are more covalent than those in $NbOCl_2$. The trend for $NbOBr_2$ falls between those of $NbOI_2$ and $NbOCl_2$ and is omitted here.

**Table S13 | Dynamical charge ($Z^*_{m^x,1}$) and $\frac{\partial u_{m^x}}{\partial \eta_1}$ of each atom in $NbOX_2$.** $m^x$ refers to the x component of each atomic displacement. We see that the values of $Z^*_{m^x,1}$ and $\frac{\partial u_{m^x}}{\partial \eta_1}$ have the same sign, contributing to a large value in the sum for $e^{ion}_{11}$. Note that $m^x$ refers to the $x$-component of each atomic displacement.

| | $Z^*_{m^x,1}$ (e) | | | $\frac{\partial u_{m^x}}{\partial \eta_1}$ (Å) | | |
|---|---|---|---|---|---|---|
| | $NbOCl_2$ | $NbOBr_2$ | $NbOI_2$ | $NbOCl_2$ | $NbOBr_2$ | $NbOI_2$ |
| Nb1 | 7.550 | 7.713 | 8.081 | 2.109 | 2.354 | 3.023 |
| Nb2 | 7.550 | 7.713 | 8.081 | 2.109 | 2.354 | 3.023 |
| X1 | -0.484 | -0.381 | -0.218 | -0.933 | -1.081 | -1.410 |
| X2 | -0.382 | -0.274 | -0.084 | -0.429 | -0.558 | -0.844 |
| X3 | -0.382 | -0.274 | -0.084 | -0.429 | -0.558 | -0.844 |
| X4 | -0.484 | -0.381 | -0.218 | -0.933 | -1.081 | -1.410 |
| O1 | -6.683 | -7.058 | -7.779 | -0.747 | -0.714 | -0.769 |
| O2 | -6.683 | -7.058 | -7.779 | -0.747 | -0.714 | -0.769 |

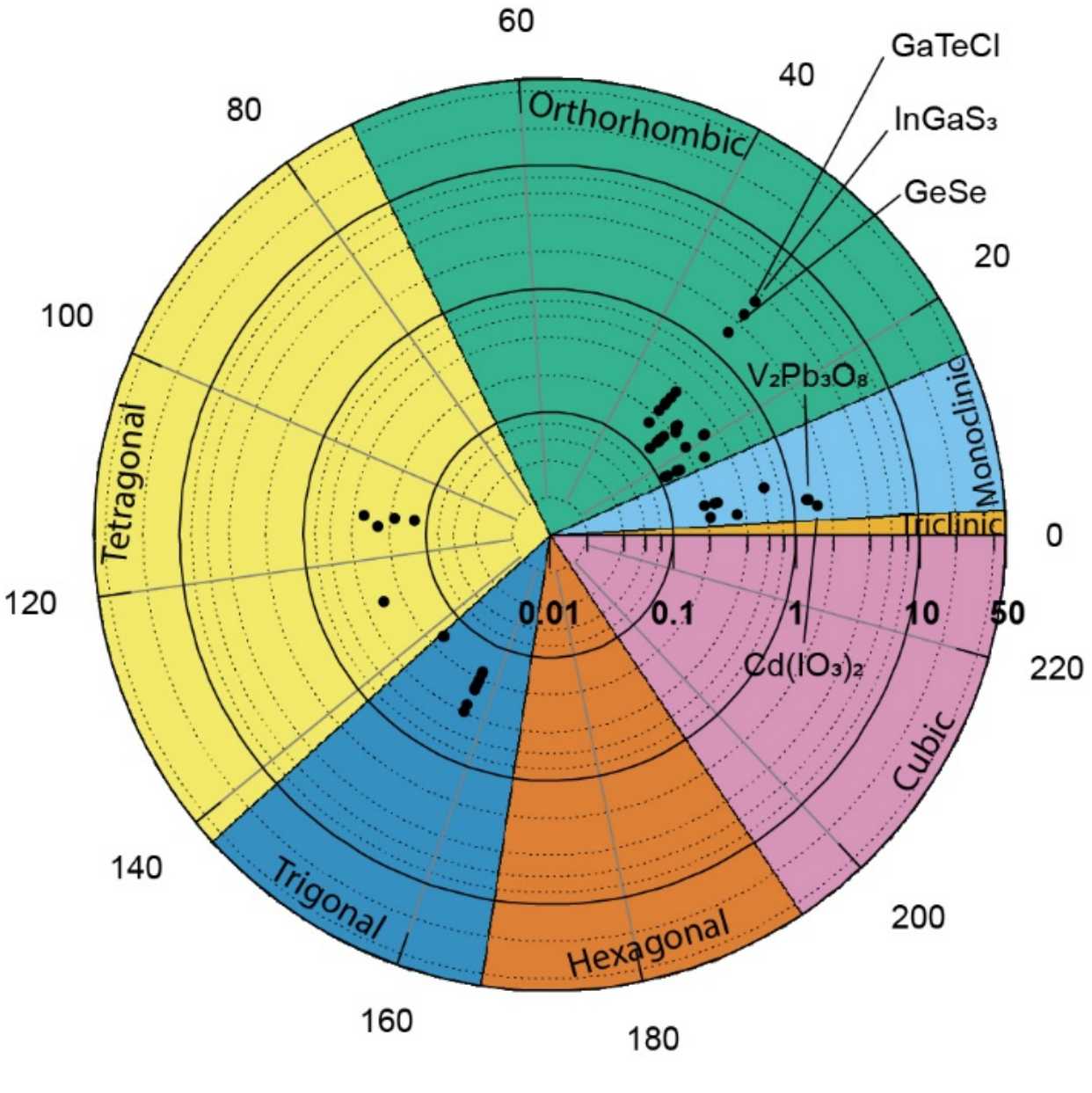

**Fig. S17 | High-throughput calculation results for maximum out-of-plane sheet piezoelectric tensor elements.** The radial axis represents the magnitude of $e_{3j}$ in units of $10^{-10}$ C $m^{-1}$ on a log scale and the angular axis represents the 230 space groups.

**Table S14 | Materials with large $e_{3j}$.**

| Material | max sheet $e_{3j}$ ($10^{-10}$ C m$^{-1}$) | |
|---|---|---|
| $InGaS_3$ | 3.349 | $e_{34}$ |
| GaTeCl | 2.429 | $e_{34}$ |
| GeSe | 1.557 | $e_{34}$ |
| $Cd(IO_3)_2$ | 1.546 | $e_{35}$ |
| $V_2Pb_3O_8$ | 1.330 | $e_{34}$ |

# II. Supplementary Discussions

## Formalisms

In this section, $\boldsymbol{e}_{ij}$ refers to the three-dimensional (3D) relaxed-ion piezoelectric stress tensor, which is defined by the relation[29]

$$\boldsymbol{e}_{ij} = \left(\frac{\partial \boldsymbol{P}_i}{\partial \boldsymbol{\eta}_j}\right)\bigg|_{\varepsilon} = -\left(\frac{\partial \boldsymbol{\sigma}_i}{\partial \boldsymbol{\epsilon}_j}\right)\bigg|_{\boldsymbol{\eta}} \qquad 1$$

where $\boldsymbol{P}$ is the electric polarization, $\boldsymbol{\eta}$ is the homogeneous strain, $\boldsymbol{\sigma}$ is the mechanical stress, and $\boldsymbol{\epsilon}$ is the homogeneous electric field. $i = \{x, y, z\}$) and $j = \{1 \ldots 6\}$ as in Voigt notation.

$\boldsymbol{e}_{ij}$ consists of two parts: clamped-ion contributions ($\boldsymbol{e}_{ij}^{el}$) as well as ionic contributions ($\boldsymbol{e}_{ij}^{ion}$)[30].

$$\boldsymbol{e}_{ij} = \boldsymbol{e}_{ij}^{el} + \boldsymbol{e}_{ij}^{ion} \qquad 2$$

$\boldsymbol{e}_{ij}^{el}$ is a second-derivative response function tensor of energy ($E$) with respect to homogeneous electric field and homogeneous strain.

$$\boldsymbol{e}_{ij}^{el} = -\frac{\partial^2 E}{\partial \epsilon_i \partial \boldsymbol{\eta}_j}\bigg|_{u} \qquad 3$$

It is considered a clamped-ion quantity because the ionic coordinates ($\boldsymbol{u}$) are not relaxed when the homogeneous electric field and strain are applied, hence it represents the piezoelectric contributions from the electrons alone.

On the other hand, $\boldsymbol{e}_{ij}^{ion}$ is defined, in an implied sum notation, as

$$\boldsymbol{e}_{ij}^{ion} = \frac{1}{\Omega_0} \boldsymbol{Z}_{mi}^{*} \frac{\partial \boldsymbol{u}_m}{\partial \boldsymbol{\eta}_j} \qquad 4$$

where $\Omega_0$ is the cell volume before deformation and $\boldsymbol{Z}_{mi}^{*}$ is the Born effective charge ($m$ is a composite label for atom and displacement directions, ranging from 1 to 3N). $\boldsymbol{e}_{ij}^{ion}$ represents the ionic contribution to $\boldsymbol{e}_{ij}$ due to the relaxation of ionic positions after the application of strain.

$\boldsymbol{e}_{ij}$ is computed using density functional perturbation theory (DFPT)[30-32] where $\boldsymbol{e}_{ij}^{ion}$ is computed in terms of the pseudo inverse of force constant matrix ($\boldsymbol{K}_{mn}$) and the internal strain tensor ($\boldsymbol{\Lambda}_{nj}$) as shown in Equation S5) [30].

$$\boldsymbol{e}_{ij}^{ion} = \frac{1}{\Omega_0} \boldsymbol{Z}_{mi}^{*} (\boldsymbol{K}^{-1})_{mn} \boldsymbol{\Lambda}_{nj} \qquad 5$$

The piezoelectric strain tensor elements ($\boldsymbol{d}_{ij}$), frequently used in experimental studies, is defined by the relation[29]

$$\boldsymbol{d}_{ij} = \left( \frac{\partial \boldsymbol{P}_i}{\partial \boldsymbol{\sigma}_j} \right) \bigg|_{\varepsilon} = - \left( \frac{\partial \boldsymbol{\eta}_i}{\partial \boldsymbol{\epsilon}_j} \right) \bigg|_{\sigma} \qquad 6$$

$\boldsymbol{d}_{ij}$ can be obtained from $\boldsymbol{e}_{ij}$ and elastic compliance tensor ($\boldsymbol{S}_{ij}$) through the following relationship

$$\boldsymbol{d}_{ij} = \boldsymbol{e}_{ik} \boldsymbol{S}_{kj} \qquad 7$$

The sheet $\boldsymbol{e}_{ij}$ defined in the main text is obtained by multiplying the 3D $\boldsymbol{e}_{ij}$ by the cell height.

The sheet $\boldsymbol{S}_{ij}$ defined in **Table S7** is obtained from 3D $\boldsymbol{S}_{ij}$ by setting elements related to the $z$-direction to 0 and dividing the rest of the elements by cell height[33].

Using both sheet $\boldsymbol{e}_{ij}$ and sheet $\boldsymbol{S}_{ij}$ for calculation of $\boldsymbol{d}_{ij}$ is equivalent to using their bulk counterparts because the scaling by height in both terms are cancelled off.

Note that when strain and electric field are simultaneously present, a more accurate formulation, as presented by Wu et al.[30], needs to be invoked. For the sake of readability, we present the "improper" formulation here while understanding that the "proper" terms as presented by Wu et al. are used in the DFPT calculations. Also, we note that the "improper" and "proper" terms are equivalent when $j = \{1, 2, 3\}$.

## Supplementary Note 1: Polarization switching in $NbOX_2$

We study the polarization switching in bulk $NbOI_2$ microscopically via polarization *versus* electric field (P–E) measurements and locally through spectroscopic PFM characterizations. The P–E ferroelectric hysteresis loop of $NbOI_2$ when an external electric field is applied to its polar axis is displayed in **Fig. S18a**. The polarization increases linearly with the field strength at low electrical fields, however, when the applied electric field is in the vicinity of the coercive field $E_c$ (~ 8.5 kV/cm), the polarization shows a drastic variation due to domain reversal; the field is large enough to switch domains with the unfavorable direction of polarization. Polarization reversal is a consequence of the motion of domain walls under the influence of strong applied fields. The process involves the redistribution of the volumes of energetically favorable and unfavorable domains. The ferroelectric domain wall motion and domain switching account for the hysteretic behavior of polarization. If the applied field strength slowly decreases, some domains would back-switch. At the zero-field point, the polarization is nonzero. The crystal reaches a zero-polarization state at the opposite $E_c$. Further increase of the field in the negative direction induces polarization switching in the opposite direction. The rounding of the hysteresis loop is ascribed to the small bandgap of $NbOI_2$ and leakage-related issues.

Scanning tip-induced switching events were recorded using spectroscopic PFM to further confirm the switching characteristics of the ultrathin $NbOX_2$ flakes. The local hysteresis curves of $NbOI_2$ and $NbOCl_2$ are shown in **Fig. S18b** and **Fig. S18c**, respectively. The phase-electric field hysteresis loop elucidates the local polarization behavior while the amplitude-electric field hysteresis loop defines the local strain response. We find that the phase can be switched by 180 ° at + 2.5 V and switched back at - 3 V, and the amplitude response displays a butterfly-like hysteric loop with dips at the same voltages as the phase curve. Meanwhile, no obvious change in the surface morphology of the $NbOX_2$ nanoflakes is found during the field cycling. The pronounced polarization reversibility at the positive and negative coercive field points and the corresponding butterfly strain-electric field hysteresis, affirm the polarization switching in ultrathin $NbOX_2$.

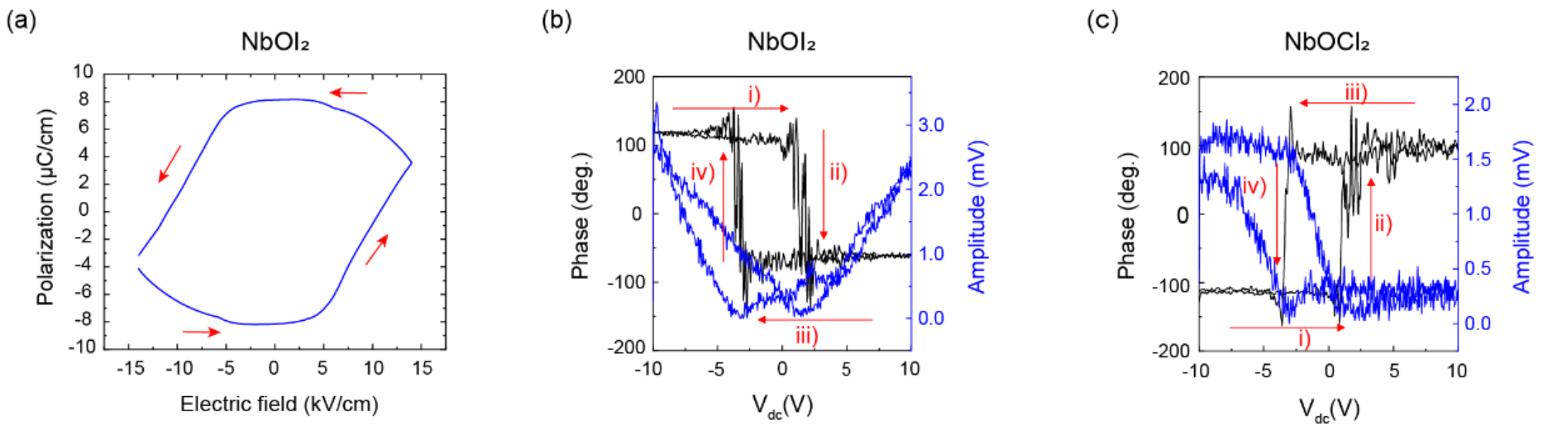

**Fig. S18 | Polar switching of $NbOX_2$.** (a) Spontaneous polarization *versus* electric field ($P$-$E$) hysteresis loop of bulk $NbOI_2$ sheets at room temperature. (b,c) Spectroscopic PFM switching loops of exfoliated (a) 4.1-nm-thick $NbOI_2$ and (b) 7.8-nm-thick $NbOCl_2$.

## Supplementary Note 2: Ferroelectric-paraelectric phase transition in $NbOI_2$

The ferroelectric-paraelectric phase transition in $NbOI_2$ was confirmed by temperature-dependent differential scanning calorimetry (DSC) and second harmonic generation (SHG) measurements. Differential scanning calorimetry (DSC) is a thermo-analytical technique that measures physical and chemical changes within a material in response to temperature. From the DSC heat flow curve, the changes in heat capacity that occurs around the ferroelectric-paraelectric phase transition can be identified as an exothermic or endothermic peak on the low-temperature side of the material's melting/sublimation peak. **Fig. S19a** depicts the DSC result of $NbOI_2$ during a heating cycle. An endothermic peak with a peak value of ~ 189.27 °C is observed in the DSC heat flow curve and assigned to the phase transition Curie temperature ($T_c$) of $NbOI_2$. This corresponds to a transition from ferroelectric phase with non-centrosymmetric C2 (No. 5) symmetry to paraelectric (PE) phase with centrosymmetric C2/m (No. 12) symmetry. SHG is highly sensitive to the inversion-symmetry breaking that accompanies a ferroelectric-paraelectric phase transition; only non-centrosymmetric structures are capable of emitting SHG light. We found that the SHG signal vanishes once $T_c$ is exceeded and emerges again upon cooling (**Fig. S19b**).

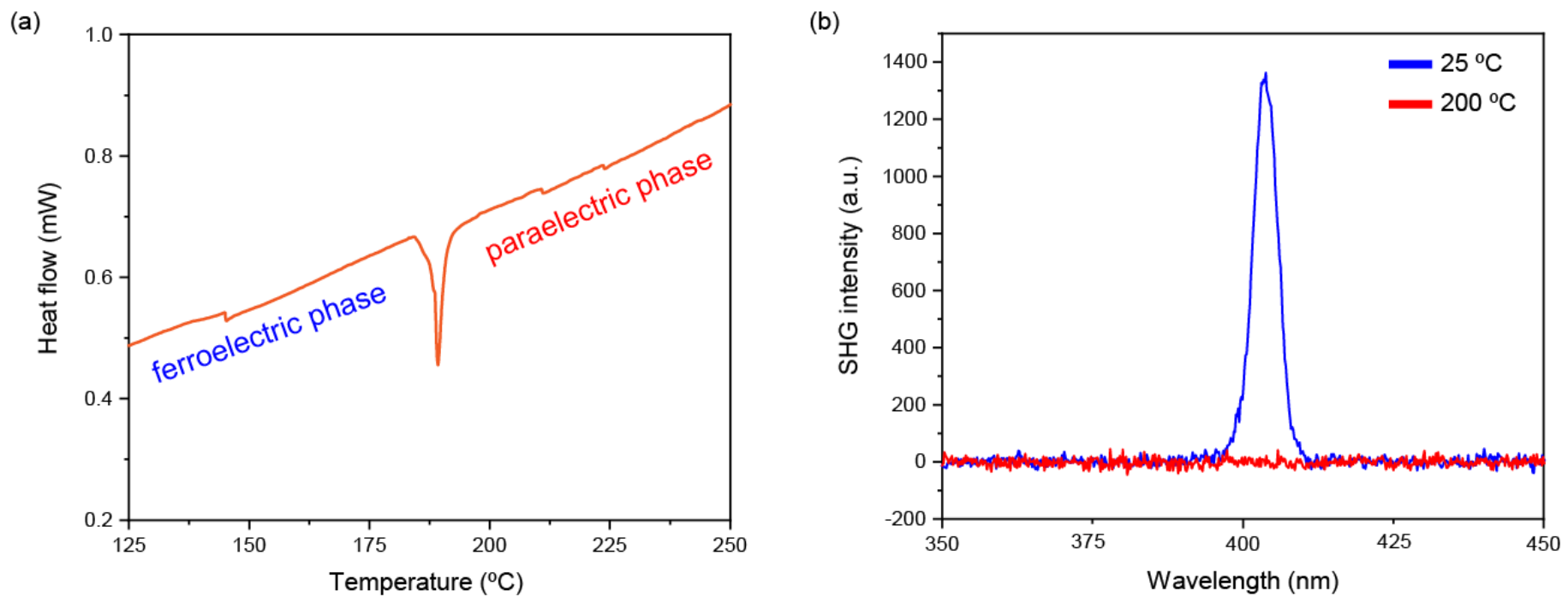

**Fig. S19 | Temperature-dependent properties of $NbOI_2$.** (a) DSC curve. (b) SHG spectra at 25 °C and 200 °C for the excitation wavelengths of $\lambda_{pump}$ = 800 nm.

## Supplementary Note 3: Pseudo-Jahn-Teller Effect

To understand the pseudo-Jahn-Teller effect (PJTE) driving the off-centre displacements of Nb in $NbOX_2$, we construct the centred symmetric structures, in which the atoms of $NbOX_2$ are relaxed with the Nb atoms constrained at $\delta x = 0$ Å, and study the valence and conduction band eigenstates. Comparing the orbital characters of these eigenstates with those of the equilibrium structures, we identify pairs of valence and conduction band states in the equilibrium structure, that are linear combinations of pairs of valence and conduction band states in the symmetric structure. This mixing of valence and conduction band states leads to an increased covalency as well as increased energy difference within each pair of states, and is accompanied by a spontaneous symmetry-breaking distortion[34] as observed here. The band indices of these valence and conduction band pairs (VB1, CB1 and VB2, CB2) are provided in **Fig. S20** below. These bands mainly comprise O $p_y$, Nb $d_{xy}$, O $p_z$ and Nb $d_{xz}$ orbitals respectively. In both VB-CB pairs, the π-like interaction between these orbitals corroborates Wheeler et al.'s conclusion that for metals with a low $d$ electron count, PJTE mixing of the metal $d_\pi$ and X $p_\pi$ orbitals favours asymmetric X-M-X bridges[35].

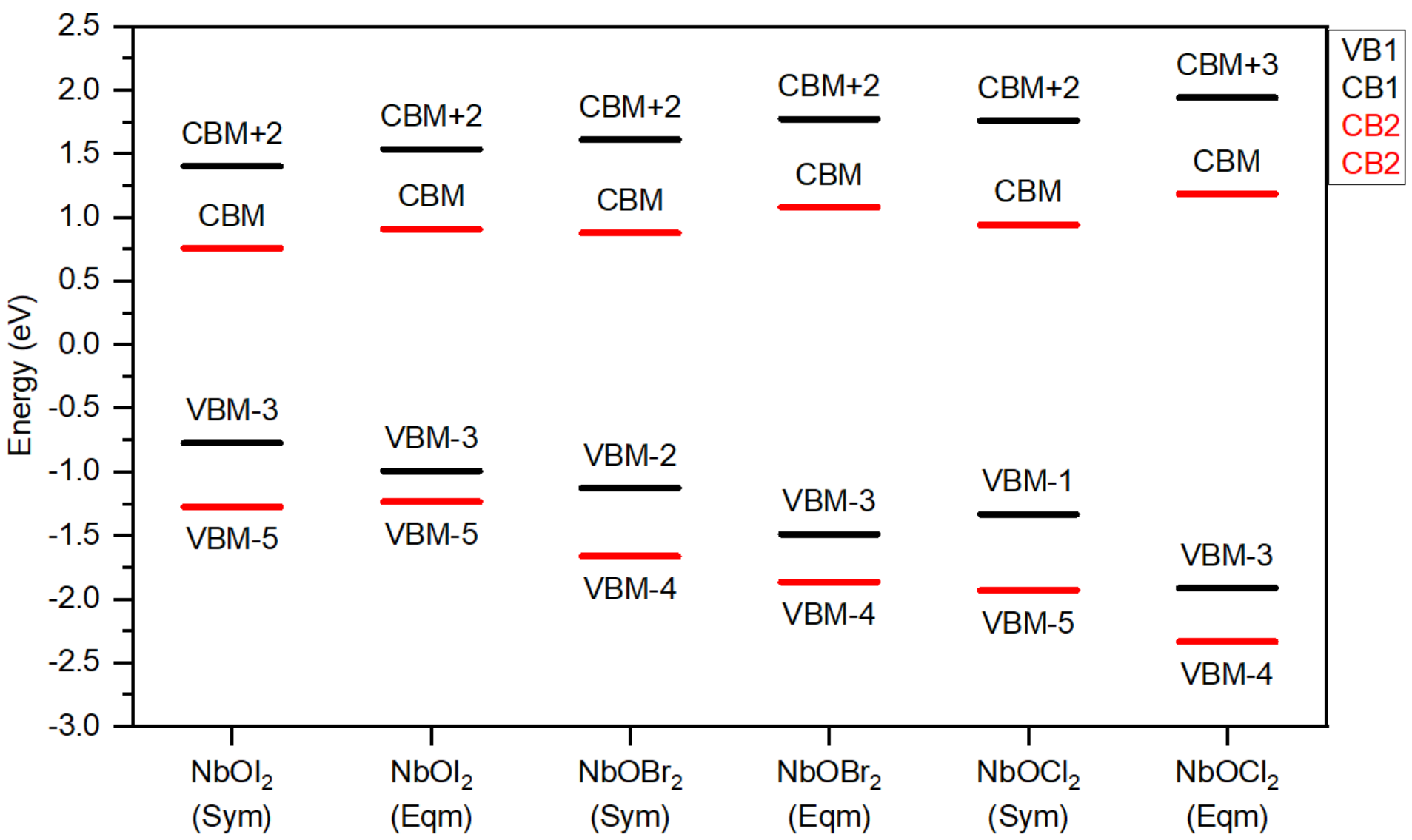

**Fig. S20 | Valence and conduction band pairs involved in inducing the symmetry-breaking distortion though the PJTE.** Band indices are provided at the Gamma point.

## Supplementary Note 4: Strain-assisted Ferroelectric Switching

As discussed in the main text, the magnitude of the applied electric field can be much reduced if one applies compressive strain to the materials. This effect is especially large for $NbOI_2$ which has the largest piezoelectric effect.

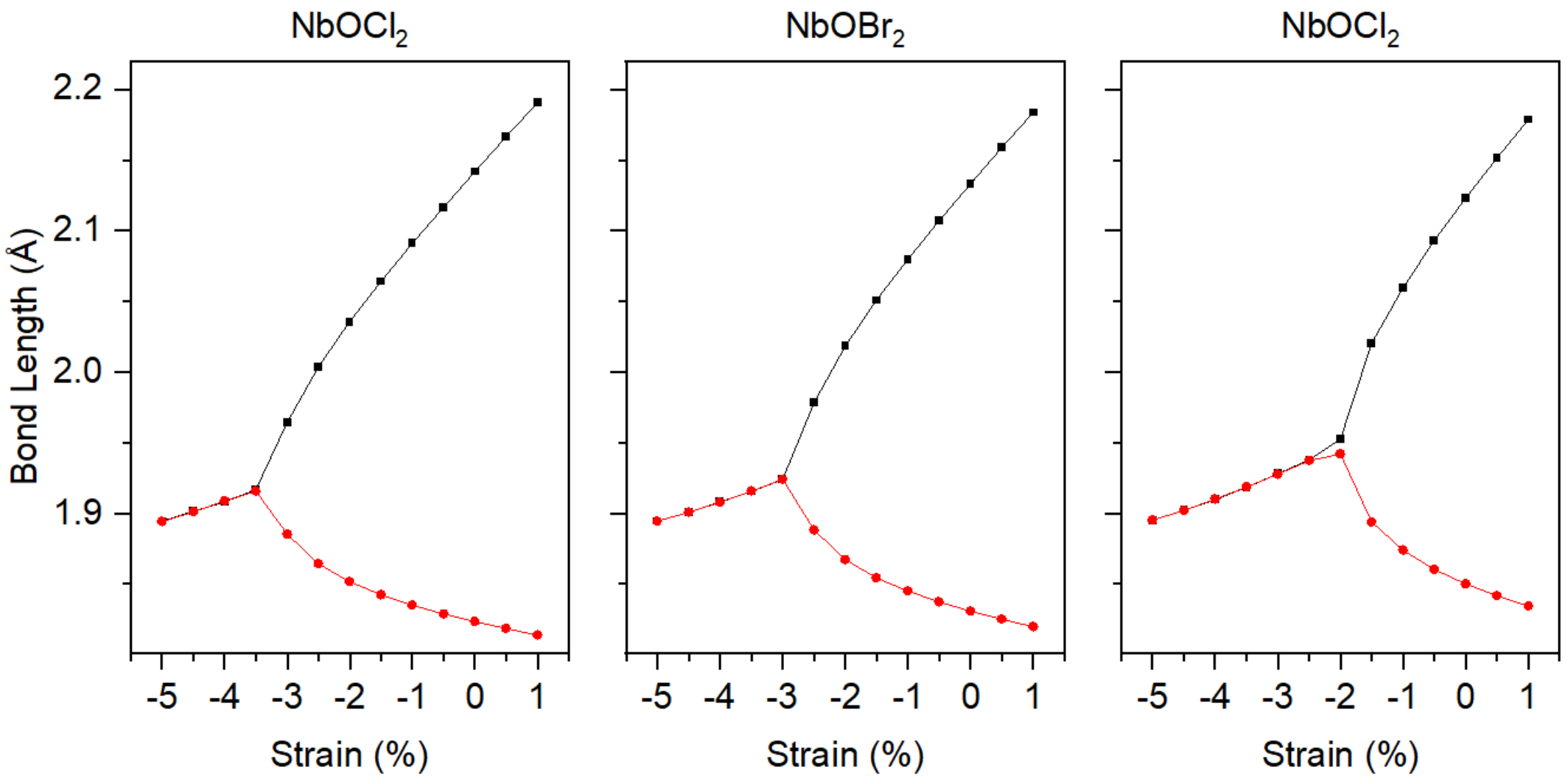

**Fig. S21 | Effect of strain on the Nb-O bond lengths in $NbOX_2$.** A centred symmetric structure is obtained when compressive strain of over -3.5%, -3.0% and -2.5% along the $x$-direction is applied to $NbOCl_2$, $NbOBr_2$ and $NbOI_2$ respectively.

The large response of Nb atoms in response to strain in the $x$-direction also enables both the magnitude of polarization and the ferroelectric switching barriers to be controlled by strain. A modest compressive strain of -2.5% in $NbOI_2$ results in a centred symmetric structure (**Supplementary Fig. S21**), allowing the polarization direction to be set with an electric field of small magnitude. The polarization magnitude can then be enhanced by applying tensile strain.

# III. High Throughput Calculation Results

**Table S15 | Table of quantities obtained from high throughput calculation workflow. Formula Pretty** is the human readable chemical formula of the material. **S.G. No.** is the space group number of the material. Non-ferroelectric **point groups**[36] are marked in bold. **Sheet plane vector direction** presents the direction of the 2D sheet's normal vector. **Height** presents the height (Å) of the unit cell. **Layer thickness** is the distance (Å), perpendicular to the 2D sheet, between the topmost and bottom most atoms. **Band gap** is in the unit of (eV). **Max abs piezo** is the magnitude of the largest piezoelectric stress tensor element in $C\ m^{-2}$. **Max abs sheet piezo** is the magnitude of the largest piezoelectric stress tensor element in $10^{-10}\ C\ m^{-1}$. **Max piezo index** is the index of the piezoelectric stress tensor element that has the largest magnitude. **Oop piezo** is a boolean reflecting if the material has a piezoelectric stress tensor element corresponding to the out-of-plane direction that has a magnitude larger than 0.005 $C\ m^{-2}$. **Max abs oop piezo**, **max abs sheet oop piezo** and **max oop piezo index** are out-of-plane counterparts of Max abs piezo, max abs sheet piezo and max piezo index respectively.

| formula pretty | 2DMatpedia ID | S.G. No. | point group | sheet plane vector direction | height | layer thickness | band gap | max abs piezo | max abs sheet piezo | max piezo index | oop piezo | max abs oop piezo | max abs sheet oop piezo | max oop piezo index |
|---|---|---|---|---|---|---|---|---|---|---|---|---|---|---|
| $SbF_3$ | 2dm-3709 | 31 | mm2 | z | 23.009 | 2.882 | 4.441 | TRUE | 2.002 | 46.072 | yxy | FALSE | | |
| $NbI_2O$ | 2dm-4281 | 25 | mm2 | z | 24.170 | 4.603 | 0.971 | TRUE | 1.279 | 30.922 | xxx | FALSE | | |
| $NbBr_2O$ | 2dm-4734 | 25 | mm2 | z | 23.866 | 4.219 | 0.917 | TRUE | 1.103 | 26.329 | xxx | FALSE | | |
| $NbCl_2O$ | 2dm-3054 | 25 | mm2 | z | 23.541 | 3.915 | 0.922 | TRUE | 1.061 | 24.974 | xxx | FALSE | | |
| $Mo(BrO)_2$ | 2dm-3188 | 26 | mm2 | z | 24.536 | 4.862 | 1.631 | TRUE | 0.678 | 16.626 | yyy | FALSE | | |
| GeSe | 2dm-4478 | 31 | mm2 | z | 23.325 | 2.599 | 1.222 | TRUE | 0.526 | 12.270 | yyy | TRUE | 0.067 | 1.557 |
| $SbAsO_3$ | 2dm-3790 | 7 | m | z | 22.588 | 4.178 | 4.040 | TRUE | 0.469 | 10.594 | xxx | TRUE | 0.010 | 0.232 |
| $InCu(PSe_3)_2$ | 2dm-3689 | 149 | **32** | z | 23.339 | 3.424 | 0.503 | TRUE | 0.435 | 10.141 | xxy | FALSE | | |
| $V_2Pb_3O_8$ | 2dm-3693 | 5 | 2 | z | 23.936 | 4.927 | 3.283 | TRUE | 0.399 | 9.539 | yyy | TRUE | 0.056 | 1.330 |
| $Zn(BH_4)_2$ | 2dm-4877 | 26 | mm2 | z | 26.663 | 6.255 | 4.499 | TRUE | 0.351 | 9.361 | xxy | FALSE | | |
| $InSnCl_3$ | 2dm-4964 | 8 | m | z | 21.349 | 2.516 | 2.959 | TRUE | 0.344 | 7.349 | xxx | FALSE | | |
| $Sn_2IF_3$ | 2dm-3802 | 59 | **mmm** | z | 24.570 | 6.755 | 3.092 | TRUE | 0.293 | 7.203 | yzx | FALSE | | |
| $NbTlBr_4O$ | 2dm-3972 | 25 | mm2 | z | 22.931 | 3.631 | 1.436 | TRUE | 0.293 | 6.708 | xxx | FALSE | | |
| $SbTeClO_3$ | 2dm-4969 | 31 | mm2 | z | 23.441 | 4.092 | 3.210 | TRUE | 0.215 | 5.037 | yxx | TRUE | 0.011 | 0.260 |
| $As_2O_3$ | 2dm-3779 | 7 | m | z | 22.204 | 3.958 | 4.294 | TRUE | 0.216 | 4.802 | xxx | TRUE | 0.011 | 0.246 |
| $Te_2Mo$ | 2dm-5370 | 187 | **-6m2** | z | 23.824 | 3.617 | 1.160 | TRUE | 0.194 | 4.617 | xxy | FALSE | | |
| $SrH_2O_3$ | 2dm-4166 | 26 | mm2 | z | 20.896 | 3.923 | 3.054 | TRUE | 0.216 | 4.515 | yyy | TRUE | 0.006 | 0.121 |
| $TaPbF_7$ | 2dm-5388 | 6 | m | z | 23.050 | 3.270 | 5.255 | TRUE | 0.187 | 4.309 | yyy | FALSE | | |
| $BaH_2O_3$ | 2dm-4554 | 26 | mm2 | z | 25.441 | 3.901 | 3.047 | TRUE | 0.166 | 4.227 | yyy | TRUE | 0.006 | 0.148 |
| $ZnH_2SeO_4$ | 2dm-3789 | 7 | m | z | 23.395 | 3.932 | 4.458 | TRUE | 0.177 | 4.131 | xyy | FALSE | | |
| $MoSe_2$ | 2dm-3409 | 187 | **-6m2** | z | 23.506 | 3.339 | 1.450 | TRUE | 0.163 | 3.839 | xxy | FALSE | | |
| $MoS_2$ | 2dm-3150 | 187 | **-6m2** | z | 23.332 | 3.122 | 1.722 | TRUE | 0.159 | 3.719 | yyy | FALSE | | |
| $TlBS_3$ | 2dm-3554 | 7 | m | z | 23.128 | 3.691 | 2.180 | TRUE | 0.153 | 3.537 | yyy | TRUE | 0.008 | 0.193 |
| $Hg_3SeO_6$ | 2dm-5414 | 8 | m | z | 21.942 | 2.188 | 1.095 | TRUE | 0.156 | 3.431 | yyy | TRUE | 0.028 | 0.609 |
| $InGaS_3$ | 2dm-3760 | 31 | mm2 | z | 26.993 | 7.216 | 1.986 | TRUE | 0.124 | 3.349 | zyz | TRUE | 0.124 | 3.349 |
| $Te_2W$ | 2dm-3050 | 187 | **-6m2** | z | 23.749 | 3.630 | 1.193 | TRUE | 0.134 | 3.187 | yxx | FALSE | | |
| CuHgSeCl | 2dm-5612 | 26 | mm2 | z | 28.830 | 3.980 | 0.798 | TRUE | 0.110 | 3.181 | yyy | TRUE | 0.008 | 0.229 |
| $As_2Se_3$ | 2dm-4755 | 31 | mm2 | x | 22.455 | 2.906 | 1.770 | TRUE | 0.141 | 3.158 | yyy | FALSE | | |
| $WSe_2$ | 2dm-3594 | 187 | **-6m2** | z | 23.485 | 3.360 | 1.543 | TRUE | 0.111 | 2.596 | xxy | FALSE | | |
| $WS_2$ | 2dm-3749 | 187 | **-6m2** | z | 23.220 | 3.141 | 1.805 | TRUE | 0.109 | 2.541 | yxx | FALSE | | |
| $AgBi(PSe_3)_2$ | 2dm-5567 | 143 | 3 | z | 23.398 | 3.582 | 1.442 | TRUE | 0.108 | 2.523 | yxx | TRUE | 0.006 | 0.150 |
| GaTeCl | 2dm-3523 | 31 | mm2 | z | 24.660 | 5.327 | 2.296 | TRUE | 0.098 | 2.429 | zyz | TRUE | 0.098 | 2.429 |
| $SrH_4O_3$ | 2dm-3672 | 26 | mm2 | z | 24.443 | 6.567 | 4.365 | TRUE | 0.094 | 2.310 | yyy | TRUE | 0.010 | 0.235 |
| $SiAs_2$ | 2dm-5490 | 26 | mm2 | z | 25.682 | 6.012 | 1.443 | TRUE | 0.089 | 2.283 | yyy | TRUE | 0.006 | 0.154 |
| $HfGeTe_4$ | 2dm-4668 | 31 | mm2 | z | 27.410 | 7.433 | 0.769 | TRUE | 0.078 | 2.128 | yyy | FALSE | | |
| $GeAs_2$ | 2dm-3619 | 26 | mm2 | z | 25.907 | 6.056 | 1.237 | TRUE | 0.080 | 2.073 | yyy | TRUE | 0.007 | 0.170 |
| $CuBi(PSe_3)_2$ | 2dm-4194 | 143 | 3 | z | 23.507 | 3.621 | 1.245 | TRUE | 0.085 | 1.994 | yxx | TRUE | 0.007 | 0.153 |
| $P_2O_5$ | 2dm-3519 | 31 | mm2 | z | 25.403 | 5.462 | 5.080 | TRUE | 0.077 | 1.967 | yyy | TRUE | 0.012 | 0.311 |
| $HgINO_3$ | 2dm-3984 | 26 | mm2 | z | 21.792 | 3.692 | 2.177 | TRUE | 0.088 | 1.916 | yxx | TRUE | 0.007 | 0.142 |
| BiTeI | 2dm-3590 | 156 | 3m | z | 23.653 | 3.773 | 1.512 | TRUE | 0.080 | 1.901 | yyy | TRUE | 0.007 | 0.177 |
| $AlHO_2$ | 2dm-4724 | 31 | mm2 | z | 25.476 | 5.350 | 4.166 | TRUE | 0.073 | 1.872 | yyy | TRUE | 0.014 | 0.354 |
| GaS | 2dm-3608 | 187 | **-6m2** | z | 24.524 | 4.643 | 2.396 | TRUE | 0.076 | 1.867 | yyy | FALSE | | |
| $Ag_3SI$ | 2dm-5200 | 4 | 2 | z | 29.013 | 9.639 | 0.547 | TRUE | 0.063 | 1.839 | yyy | TRUE | 0.012 | 0.339 |
| $NaTaCl_6$ | 2dm-3691 | 4 | 2 | z | 26.982 | 9.434 | 2.919 | TRUE | 0.068 | 1.821 | yyy | TRUE | 0.008 | 0.206 |
| $ZrGeTe_4$ | 2dm-3329 | 31 | mm2 | z | 27.565 | 7.481 | 0.689 | TRUE | 0.066 | 1.821 | yyy | TRUE | 0.010 | 0.272 |
| GaSe | 2dm-3530 | 187 | **-6m2** | z | 24.648 | 4.822 | 1.790 | TRUE | 0.073 | 1.789 | yyy | FALSE | | |
| $InAg(PSe_3)_2$ | 2dm-3598 | 149 | **32** | z | 23.469 | 3.563 | 0.901 | TRUE | 0.076 | 1.785 | yxx | FALSE | | |
| $AsPO_4$ | 2dm-5240 | 31 | mm2 | z | 25.424 | 4.837 | 4.364 | TRUE | 0.068 | 1.726 | yxx | FALSE | | |
| $As_2S_3$ | 2dm-4821 | 31 | mm2 | x | 21.848 | 2.594 | 2.299 | TRUE | 0.079 | 1.723 | yyy | FALSE | | |
| $UCO_5$ | 2dm-4524 | 25 | mm2 | z | 23.359 | 3.571 | 2.294 | TRUE | 0.070 | 1.638 | xxx | FALSE | | |
| $SiP_2$ | 2dm-4912 | 26 | mm2 | z | 25.623 | 5.592 | 1.550 | TRUE | 0.063 | 1.609 | yyy | FALSE | | |
| $VAg(PSe_3)_2$ | 2dm-4708 | 5 | 2 | z | 23.438 | 3.548 | 0.330 | TRUE | 0.066 | 1.556 | xxy | FALSE | | |
| $Cd(IO_3)_2$ | 2dm-3721 | 4 | 2 | z | 26.236 | 7.374 | 3.531 | TRUE | 0.059 | 1.546 | zzx | TRUE | 0.059 | 1.546 |
| $B_2S_2O_9$ | 2dm-3130 | 5 | 2 | z | 26.564 | 7.282 | 6.834 | TRUE | 0.058 | 1.543 | xxx | TRUE | 0.048 | 1.273 |
| $Ca(AuF_6)_2$ | 2dm-3472 | 115 | **-42m** | z | 23.867 | 7.078 | 1.639 | TRUE | 0.063 | 1.506 | xzx | FALSE | | |
| BN | 2dm-4991 | 187 | **-6m2** | z | 19.935 | 0.000 | 4.711 | TRUE | 0.073 | 1.462 | xxy | FALSE | | |
| $Mn(CuCl_2)_2$ | 2dm-5365 | 25 | mm2 | z | 20.365 | 2.840 | 0.926 | TRUE | 0.071 | 1.455 | xzx | TRUE | 0.010 | 0.213 |
| $GaAg(PSe_3)_2$ | 2dm-4552 | 149 | **32** | z | 23.369 | 3.462 | 0.938 | TRUE | 0.061 | 1.418 | yxx | FALSE | | |
| $Hg_2P_2S_7$ | 2dm-3704 | 5 | 2 | z | 24.357 | 6.217 | 2.111 | TRUE | 0.056 | 1.365 | xyy | FALSE | | |
| BiTeBr | 2dm-4356 | 156 | 3m | z | 23.305 | 3.566 | 1.607 | TRUE | 0.053 | 1.241 | xxy | FALSE | | |
| $Hg_3AsS_4Cl$ | 2dm-4753 | 156 | 3m | z | 22.449 | 3.066 | 2.125 | TRUE | 0.055 | 1.233 | xxy | FALSE | | |
| $HfFeCl_6$ | 2dm-5854 | 5 | 2 | z | 21.980 | 2.939 | 0.241 | TRUE | 0.053 | 1.174 | xyy | FALSE | | |
| $Nb_3TeI_7$ | 2dm-3841 | 156 | 3m | z | 23.847 | 3.956 | 0.611 | TRUE | 0.040 | 0.945 | xxy | TRUE | 0.009 | 0.206 |
| $Hg_3AsSe_4Br$ | 2dm-4674 | 156 | 3m | z | 22.795 | 3.338 | 1.803 | TRUE | 0.040 | 0.922 | yyy | FALSE | | |

| | | | | | | | | | | | | | |
|---|---|---|---|---|---|---|---|---|---|---|---|---|---|
| $ScAg(PS_3)_2$ | 2dm-5836 | 149 | **32** | z | 23.162 | 3.358 | 2.045 | TRUE | 0.040 | 0.922 | xxy | FALSE | | |
| $InAg(PS_3)_2$ | 2dm-3602 | 149 | **32** | z | 23.185 | 3.402 | 1.365 | TRUE | 0.040 | 0.922 | yxx | FALSE | | |
| $Ta_3SeI_7$ | 2dm-5470 | 156 | 3m | z | 23.628 | 3.772 | 0.700 | TRUE | 0.037 | 0.886 | yxx | TRUE | 0.010 | 0.240 |
| $Nb_3SBr_7$ | 2dm-3765 | 156 | 3m | z | 23.159 | 3.451 | 0.801 | TRUE | 0.037 | 0.865 | xxy | TRUE | 0.011 | 0.245 |
| $Ta_3TeI_7$ | 2dm-5496 | 156 | 3m | z | 23.831 | 3.998 | 0.667 | TRUE | 0.036 | 0.863 | yxx | TRUE | 0.007 | 0.169 |
| InSe | 2dm-3459 | 187 | **-6m2** | z | 24.994 | 5.381 | 1.386 | TRUE | 0.034 | 0.848 | yyy | FALSE | | |
| $LaBr_2$ | 2dm-5867 | 187 | **-6m2** | z | 23.350 | 3.828 | 0.625 | TRUE | 0.036 | 0.829 | yxx | FALSE | | |
| BiTeCl | 2dm-3732 | 156 | 3m | z | 23.181 | 3.375 | 1.778 | TRUE | 0.035 | 0.822 | yyz | FALSE | | |
| $Nb_3TeCl_7$ | 2dm-3785 | 156 | 3m | z | 23.374 | 3.723 | 0.772 | TRUE | 0.032 | 0.751 | xxy | FALSE | | |
| $Ta_3SBr_7$ | 2dm-5348 | 156 | 3m | z | 23.177 | 3.468 | 0.872 | TRUE | 0.032 | 0.745 | yyy | TRUE | 0.009 | 0.206 |
| $CuO_2F$ | 2dm-4542 | 17 | **222** | z | 19.541 | 1.969 | 1.044 | TRUE | 0.038 | 0.736 | yzx | FALSE | | |
| $CdTeMoO_6$ | 2dm-4591 | 113 | **-42m** | z | 27.457 | 7.962 | 3.555 | TRUE | 0.026 | 0.703 | yzx | TRUE | 0.009 | 0.244 |
| CaHClO | 2dm-4557 | 156 | 3m | z | 23.101 | 3.583 | 3.598 | TRUE | 0.028 | 0.650 | yyz | TRUE | 0.014 | 0.334 |
| $ZrCl_2$ | 2dm-3706 | 187 | **-6m2** | z | 23.475 | 3.427 | 1.028 | TRUE | 0.027 | 0.639 | xxy | FALSE | | |
| $MnTeMoO_6$ | 2dm-3666 | 18 | **222** | z | 27.418 | 7.809 | 2.517 | TRUE | 0.021 | 0.571 | yzx | FALSE | | |
| $Ge_3Sb_2O_9$ | 2dm-3499 | 174 | **-6** | z | 23.773 | 4.784 | 3.950 | TRUE | 0.021 | 0.501 | yxy | FALSE | | |
| $LaHBr_2$ | 2dm-4199 | 187 | **-6m2** | z | 23.774 | 3.900 | 3.767 | TRUE | 0.019 | 0.462 | yxx | FALSE | | |
| $Hg_3(BO_3)_2$ | 2dm-4803 | 189 | **-6m2** | z | 18.639 | 0.003 | 3.401 | TRUE | 0.024 | 0.452 | xxx | FALSE | | |
| $Nb_3I_8$ | 2dm-5497 | 156 | 3m | z | 23.975 | 4.058 | 0.233 | TRUE | 0.017 | 0.418 | xxy | FALSE | | |
| $H_3BrO$ | 2dm-5037 | 156 | 3m | z | 21.265 | 1.116 | 5.225 | TRUE | 0.018 | 0.386 | zxx | TRUE | 0.018 | 0.386 |
| $TlAsO_4$ | 2dm-5146 | 111 | **-42m** | z | 22.060 | 2.087 | 1.492 | TRUE | 0.017 | 0.376 | xyz | TRUE | 0.015 | 0.322 |
| $Sn(PS_3)_2$ | 2dm-5267 | 149 | **32** | z | 22.825 | 3.369 | 1.367 | TRUE | 0.016 | 0.365 | yyy | FALSE | | |
| $Ag_2SO_4$ | 2dm-4885 | 21 | **222** | z | 20.124 | 1.791 | 2.248 | TRUE | 0.016 | 0.317 | zxy | TRUE | 0.016 | 0.317 |
| $Ag_2SeO_4$ | 2dm-3267 | 21 | **222** | z | 20.413 | 2.028 | 1.646 | TRUE | 0.015 | 0.311 | zxy | TRUE | 0.015 | 0.311 |
| $TmAg(PSe_3)_2$ | 2dm-5578 | 149 | **32** | z | 23.395 | 3.598 | 1.839 | TRUE | 0.013 | 0.302 | xxy | FALSE | | |
| $Nb_3Cl_8$ | 2dm-5206 | 156 | 3m | z | 23.184 | 3.497 | 0.246 | TRUE | 0.013 | 0.290 | yyy | FALSE | | |
| $CuSe_2Cl$ | 2dm-4225 | 17 | **222** | z | 22.517 | 2.805 | 0.926 | TRUE | 0.013 | 0.284 | yzx | TRUE | 0.007 | 0.154 |
| NaHO | 2dm-5304 | 129 | **4/mmm** | z | 23.838 | 4.605 | 2.837 | TRUE | 0.012 | 0.281 | zzx | TRUE | 0.012 | 0.281 |
| $AgO_2F$ | 2dm-4445 | 17 | **222** | z | 22.089 | 2.098 | 1.062 | TRUE | 0.012 | 0.256 | zxy | TRUE | 0.012 | 0.256 |
| $CuSe_2Br$ | 2dm-4942 | 17 | **222** | z | 22.518 | 3.126 | 0.933 | TRUE | 0.011 | 0.248 | yzx | TRUE | 0.006 | 0.144 |
| $Cu_2WS_4$ | 2dm-4517 | 111 | **-42m** | z | 22.442 | 2.601 | 1.251 | TRUE | 0.010 | 0.231 | xyz | TRUE | 0.006 | 0.125 |
| $LiBH_4$ | 2dm-3894 | 156 | 3m | z | 21.314 | 1.664 | 6.282 | TRUE | 0.010 | 0.222 | zxx | TRUE | 0.010 | 0.222 |
| $Cu_2SO_4$ | 2dm-3586 | 21 | **222** | z | 20.331 | 1.786 | 2.120 | TRUE | 0.010 | 0.208 | zxy | TRUE | 0.010 | 0.208 |
| $Cu_2WSe_4$ | 2dm-3107 | 111 | **-42m** | z | 22.514 | 2.839 | 1.233 | TRUE | 0.009 | 0.207 | xyz | TRUE | 0.008 | 0.181 |
| $LiH_2N$ | 2dm-3071 | 113 | **-42m** | z | 23.764 | 3.461 | 3.202 | TRUE | 0.008 | 0.194 | xyz | FALSE | | |
| $ScAg(PSe_3)_2$ | 2dm-5821 | 149 | **32** | z | 23.376 | 3.545 | 1.694 | TRUE | 0.008 | 0.188 | yyy | FALSE | | |
| $CuTe_2Br$ | 2dm-5156 | 17 | **222** | z | 22.242 | 3.044 | 0.912 | TRUE | 0.008 | 0.183 | yzx | TRUE | 0.005 | 0.120 |
| $CuTe_2Cl$ | 2dm-4859 | 17 | **222** | z | 21.769 | 2.749 | 0.886 | TRUE | 0.008 | 0.180 | yzx | TRUE | 0.005 | 0.112 |
| $AuBrO_2$ | 2dm-4476 | 17 | **222** | z | 21.615 | 3.040 | 0.963 | TRUE | 0.008 | 0.177 | xyz | FALSE | | |
| $AuO_2F$ | 2dm-4797 | 17 | **222** | z | 20.562 | 2.153 | 1.350 | TRUE | 0.008 | 0.169 | yzx | FALSE | | |
| $AuClO_2$ | 2dm-3945 | 17 | **222** | z | 22.425 | 2.790 | 1.088 | TRUE | 0.007 | 0.152 | xyz | FALSE | | |
| $CuTe_2I$ | 2dm-5239 | 17 | **222** | z | 22.415 | 3.385 | 0.960 | TRUE | 0.006 | 0.144 | yzx | TRUE | 0.006 | 0.143 |
| $ErAg(PSe_3)_2$ | 2dm-5480 | 149 | **32** | z | 23.481 | 3.602 | 1.819 | TRUE | 0.006 | 0.140 | xxy | FALSE | | |
| $ZnCl_2$ | 2dm-4713 | 115 | **-42m** | z | 22.100 | 2.724 | 4.247 | TRUE | 0.004 | 0.088 | zxx | FALSE | | |
| $Li_2WS_4$ | 2dm-5501 | 111 | **-42m** | z | 22.171 | 2.486 | 1.933 | TRUE | 0.003 | 0.058 | yzx | FALSE | | |